\documentclass[10pt,twocolumn]{IEEEtran}
\usepackage{subcaption}
\usepackage[table,xcdraw]{xcolor}
\usepackage{amsmath}
\usepackage{dsfont}
\usepackage{bbm}
\usepackage{amsthm}
\usepackage{changepage}
\usepackage{tabularx}
\usepackage{graphicx}
\usepackage{cite}
\usepackage{algorithm,algcompatible}

\newtheorem{re}{Remark}
\usepackage{tcolorbox}
\usepackage{graphicx,dblfloatfix}
\usepackage[noend]{algpseudocode}

\usepackage[colorlinks,bookmarksopen,bookmarksnumbered,citecolor=blue,urlcolor=blue]{hyperref}

\begin{document}
\title{E2E Delay Guarantee for the Tactile Internet via joint NFV and Radio Resource Allocation }
\author{Narges Gholipoor, Hamid Saeedi, Nader Mokari, Eduard Jorswieck
	\thanks{}
	\thanks{}
	\thanks{ }}

\maketitle

\begin{abstract} The \textcolor{black}{Tactile Internet (TI)} is one of the next generation wireless network services with end to end (E2E) delay as low as 1~ms. \textcolor{black}{Since this ultra low E2E delay cannot be met in the current 4G network architecture}, it is necessary to investigate this service in the next generation wireless network by considering new technologies such as network function virtualization (NFV). On the other hand, given the importance of E2E delay in the TI service, it is crucial to consider the delay of all parts of the network, including the radio access part and \textcolor{black}{the NFV core part}. In this paper, \textcolor{black}{for the first time,} we investigate  the joint radio resource allocation (R-RA) and NFV resource allocation (NFV-RA) in a heterogeneous network
 where
queuing delays, transmission delays, and delays resulting from virtual network function (VNF) execution are \textcolor{black}{jointly} considered. For this setup, we formulate a new resource allocation (RA) problem to minimize the total cost function subject to guaranteeing E2E delay of each user. \textcolor{black}{Since the proposed optimization problem is highly non-convex, we exploit alternative search method (ASM), successive convex approximation (SCA),  and heuristic algorithms to solve it.}
Simulation results reveal that in the proposed scheme can significantly 
reduce the network costs compared to the case where the two problems are optimized separately.
\end{abstract}


\IEEEpeerreviewmaketitle

\section{Introduction}
\subsection{\textbf{State of the Art}}
In the next generation wireless networks, \textcolor{black}{heterogeneous} services with different requirements are proposed. These services are generally classified as enhanced mobile broadband (eMBB), massive machine-type communications (mMTC), and ultra-reliable and low latency communications (URLLC). eMBB  requires high data rate while mMTC requires  high number of connections and the data rate of each connection is low. URLLC  such as the Tactile Internet (TI), requires ultra low end-to-end (E2E) delay and ultra high reliability \cite{8403963, popovski20185g}.

The {TI} requires an E2E delay of about $1$~ms to transmit the sense of touch remotely. The {TI} has many applications, including remote monitoring, remote surgery, distance education, and remote driving. Due to the importance of E2E delay in the {TI} service, it is necessary to consider all delays in the network to satisfy this requirement \cite{she2016ensuring,simsek20165g,fettweis2014tactile,steinbach2012haptic}.

Based on current practice in the networks, each of the services requires special physical equipment and  servers at base stations (BSs). Therefore, to provide a new service, it is necessary to deploy the required physical equipment (middle-boxes) such as special servers, which are not cost-effective for the operators. Moreover, providing space and energy for various middle-boxes and managing them are costly.  Network function virtualization (NFV) is a promising method to deal with these bottlenecks \cite{han2015network,mijumbi2016network,8115155}.

NFV reduces \textcolor{black}{capital expenditure (CAPEX) and operational expenditure (OPEX)} and increases the management capabilities while optimizing the utilization of network equipment and servers, e.g., middle-boxes. The purpose of NFV is to shift middle-box processing from hardware to software \cite{martins2014clickos}. In NFV, network functions (NF) are implemented virtually on the servers which are called virtual network functions (VNFs). NFV facilitates the installation and implementation of VNFs  on the various servers at any place and time in the network \cite{7534741}.

In order to deploy a network service (NS), it is necessary to pass the traffic through a set of middle-boxes (i.e. VNFs) in a specific order, each performing a particular operation. Choosing suitable middle-boxes and steering traffic between them are two important key challenges in NFV-based systems. Hence, for each NS, the number of VNFs, their chains, and the location of each VNF on the servers should be determined  \cite{martins2014clickos,7534741}. Thus, one of the challenges in NFV deployment is how to combine, locate, and schedule VNFs to \textcolor{black}{realize} an NS.

The NFV resource allocation (NFV-RA) includes three stages: 1) \underline{VNF composition} to execute an NS that is called chain composition,  2) virtual links and servers allocation which is called \underline{VNF embedding}, and 3)   \underline{VNF scheduling} to execute an NS \cite{7243304, riera2014virtual}. In many cases, the allocation of resources is done such that the execution time of NF's is minimized.  For delay-sensitive services, NFV-RA should be performed such that the amount of delay is kept less than a threshold value. Therefore, NFV-RA is very important for delay-sensitive services and it is necessary to consider all three stages \textcolor{black}{jointly}.  

In \cite{cao2017vnf,monteleone2013session,riggio2016scheduling}, the VNF placement optimization is studied by considering the maximum link utilization.  VNF placement and scheduling are investigated \textcolor{black}{jointly} in \cite{mijumbi2015design} by considering the buffer capacity of nodes and the processing time of VNFs. The authors in \cite{beck2015coordinated} study joint VNF composition and placement to minimize bandwidth utilization by keeping the overall runtime below the threshold value.    Joint optimization of three stages of NFV-RA is studied in \cite{wang2016joint} by considering network cost and service performance.

In wireless cellular networks, allocation of the radio resources such as power, bandwidth, and resource blocks should be optimized according to the service requirements needed for multiple users (e.g., data rate and delay). In practice, Radio Resource Allocation (R-RA) plays a key role for guaranteeing the quality of service (QoS). Given that the {TI} is highly sensitive to delay, R-RA  becomes even more important in this case  \cite{she2016ensuring,simsek20165g}. 
In \cite{she2016ensuring},  a cross-layer optimization is proposed for guaranteeing the QoS parameters such as queuing delay and reliability of the {TI} service.  In that work, the QoS requirement is satisfied by considering the transmit power and bandwidth constraints.  In  \cite{she7636814tactile},  both queuing delay and uplink (UL)/downlink (DL) transmission delays are considered in a system with single roadside BS. In this work, energy efficiency (EE) is maximized by considering the QoS requirements and optimal resource allocation. UL transmission design with massive machine type devices in the {TI} is studied in  \cite{she2016uplink}.  In that work, the authors investigate the effect of diversity on reliability and the QoS for the {TI} service.  A R-RA scheme for the {TI} in LTE-A network is investigated in \cite{aijaz2016towards} for the first time in which single carrier frequency division multiple access (SC-FDMA) and orthogonal frequency division multiple access (OFDMA) are considered for UL and DL transmission, respectively. \textcolor{black}{In \cite{gholipoor2018cross}, a R-RA scheme for the {TI} in the single cell network based on sparse code multiple access (SCMA) is proposed in which queuing delay both at the source and the base station is considered. In  \cite{gholipoor2019cloud,gholipoor2019resource}, a R-RA for the {TI} in the cloud radio access network (C-RAN) architecture of the next generation wireless network  is  proposed. In that work, a multi cell network assisted via power domain non-orthogonal multiple access (PD-NOMA) is considered.}

\subsection{\textbf{Our Contributions}} 
 Since the {TI} is supposed to be implemented in the next generation of wireless networks,  it is of great importance that it is studied and analyzed in the framework of the next generation of wireless network technologies. On the other hand,    
 	the {TI} is a delay-sensitive service, and hence all of the parameters that affect the E2E delay should be considered. This includes queuing delays, transmission delays, and delays resulting from the execution of NFs. 
 	This means that current frameworks in which  R-RA and NFV-RA are considered separately, may not provide an efficient allocation of the network \textcolor{black}{resources}. In other words in this case, more resources might be necessary to meet the E2E delay requirements. As such, joint R-RA and NFA-RA has to be implemented.

In this paper, we consider a joint R-RA  and NFV-RA  framework in a heterogeneous cellular network. In this system model, we consider multiple users with different types of services in which our \textcolor{black}{target} is to minimize the total cost subject to E2E delay constraints. The contributions of this paper can be summarized as follows:
	\begin{itemize}
		\item We consider and analyze the requirements of {TI} Service in an NFV-based framework for the first time.
			\item We perform a novel joint R-RA and NFV-RA for the {TI} service, referred to as the joint approach (JA). 
			\item Using the proposed framework, we show a considerable amount of \textcolor{black}{costs in term of transmit power and NFV execution time} saving  compared to the separate approach (SA).
			\item We consider a practical scenario in which each user can request a service with its own delay requirement.  In prior works, \textcolor{black}{heterogeneity of services  are not considered.}
			
				\item In this paper, we propose a general solution for placement and scheduling of VNFs by considering delay requirements and radio resource limitations.
			\end{itemize}
			\subsection{Paper Organization}
		The rest of this paper is as follows. In Section \ref{Systemmodel}, the system model and related constraints are described. In Section \ref{optimizationsolve}, we propose the joint optimization problem and its solution. In Section \ref{complexity}, computational complexity is calculated. Numerical results are presented in Section \ref{simulationresults}. Finally, Section \ref{conclusion} concludes the paper.

\section{System Model}\label{Systemmodel}
\subsection{General Description}
\begin{figure}[t]
 	\centering
 	\centerline{\includegraphics[width=0.5\textwidth]{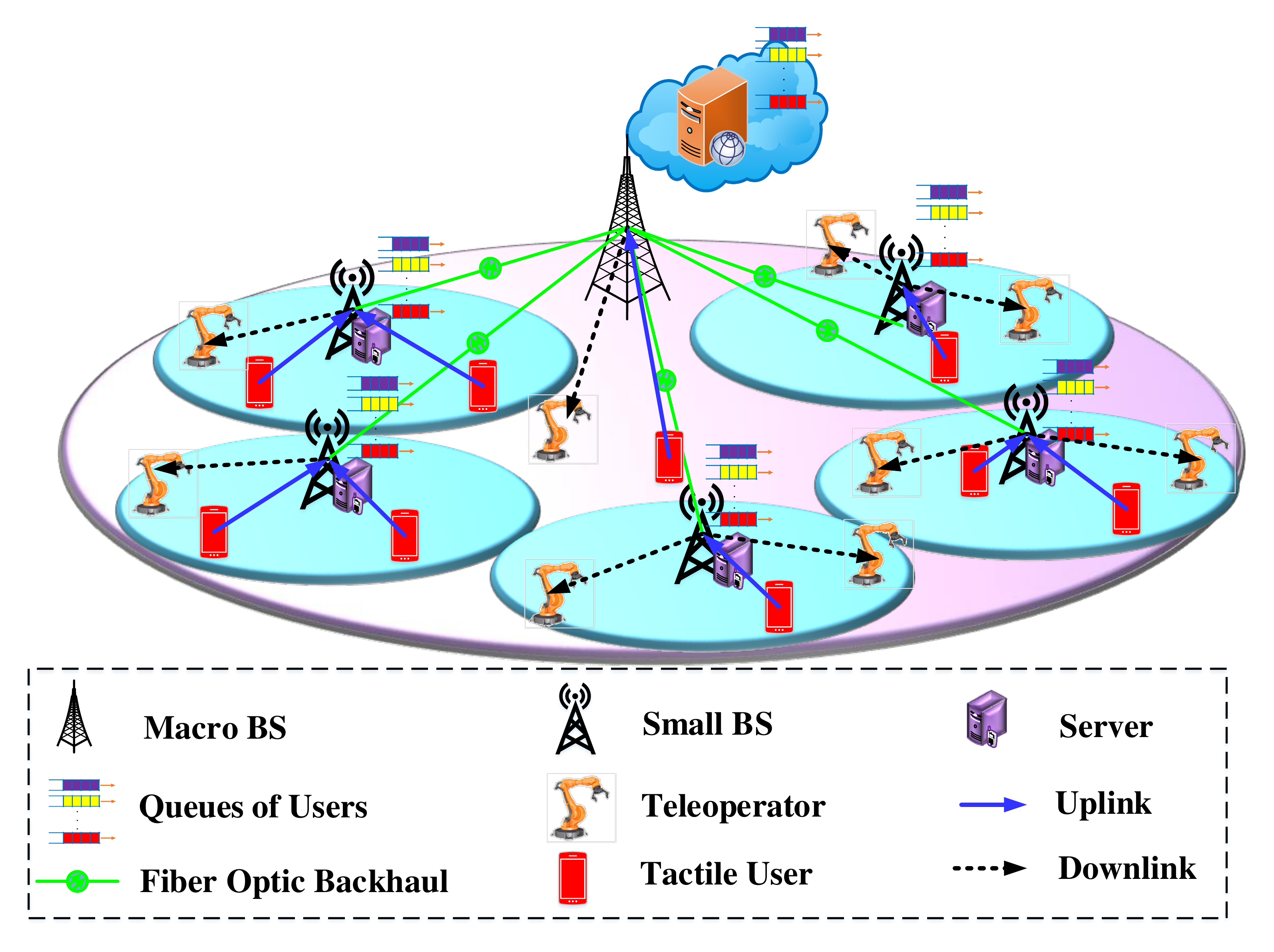}}
 	\caption{ Proposed System Model }
 	\label{pic}
 \end{figure}

We consider a heterogeneous cellular network where all BSs are connected together via  backhaul links. Furthermore, there exist multiple tactile users and teleoperators where each tactile user sends its data to its paired teleoperator\footnote{The teleoperator is a tactile data receiver such as actuator that executes a command based on received data.}, via UL and DL transmission links. As shown in Fig. \ref{pic},  we consider a heterogeneous network consisting of a macro BS (MBS) and several small BSs (SBS)  denoted by $\mathcal{J}'=\{1, ..., J\}$.  We \textcolor{black}{indicate} the MBS with index $0$.
Therefore,  $\mathcal{J}=\{0,1, ..., J\}$ is the set of all BSs. Moreover, in the proposed system model, there is a set of $\boldsymbol{\mathcal{{O}}_j}=\{1,\dots,O_j\}$ teleoperators at BS $j$ which receive the tactile data. The total number of teleoperators in our system model is equal to $\mathcal{O}_\text{Tot}=\bigcup_{j\in \mathcal{J}}\mathcal{O}_j$. 
In addition, in our system model, the set of all subcarriers for UL transmission and DL transmission are denoted by  $\boldsymbol{\mathcal{K}}=\{1,\dots,K\}$  and $\boldsymbol{\mathcal{L}}=\{1,\dots,L\}$, respectively.

It should be noted that the proposed system model works in frequency division duplexing (FDD) mode. 
 In this system model, we consider orthogonal frequency division multiple access (OFDMA) in which each subcarrier is allocated to at most one user in each time slot.
Given that NFs of one NS are located in different BSs, for executing each NS, the data should be transferred between BSs. We assume that there exist $\mathcal{S}\in \{1,...,S\}$ NSs in our considered system model.  NS $s$ consists of a set of NFs $ \mathcal{F}_s=\{1,...,F_s\}$. We assume that the chain of NFs of each NS is known and fixed and we focus on scheduling and placement of NFs. In the proposed system model,    $\mathcal{U}^s_j= \{1,...,U^s_j\}$ tactile users request service $s$ at BS $j$ and the total number of tactile users  is equal to $\mathcal{U}_ \text{Tot}=\bigcup_{j\in \mathcal{J}}\bigcup_{s\in \mathcal{S}}\mathcal{U}^s_j$. \textcolor{black}{The proposed joint R-RA and NFV-RA is an offline algorithm i.e., at the beginning of each frame after all users send their requests, joint R-RA and NFV-RA  should be performed for all received requests.  It is worth noting that we assume the channel power gain is constant in each frame and varies frame by frame. }
\begin{re}
	In this paper, we assume that the Backhaul topology is full mesh \cite{roos2006broadband,7497015,liao2013performance,4224302,8651725} and we only \textcolor{black}{consider} the capacity of the backhaul links regardless of its type (Fiber/mmWave).
\end{re}
\begin{re}
	It is assumed that the next generation wireless radio access network (RAN) is virtual enabled meaning that some RAN functions are implemented virtually including Virtualized BS \cite{7452271,6553675}. Given that the focus of this paper is on the delay-sensitive {TI} service, it is necessary that some processing functions, such as mobile edge computing (MEC) functions, are executed at each BS \cite{filippou2019flexible,8533343}. Therefore, we consider a lightweight server at each SBS and a high volume server (HVS) at the MBS \cite{8436039,filippou2019flexible,8533343}.
\end{re}
 \subsection{ \textcolor{black}{Radio Resource Allocation}}
We consider a binary variable $x_{u^s_j}^{k}$ which is set to 1 if subcarrier $k$ is assigned to tactile user $u^s_j$ at BS $j$ with requested service $s$ and otherwise it  is set to 0 , i.e.,
\begin{equation} \nonumber
\begin{split}
& x_{u^s_j}^{k}=\begin{cases}
1, &\text{If \text{subcarrier~} $k$ is assigned to tactile user $u^s_j$}\\ &   \text{~at BS~$j$~with requested NS $s$}, \\
0, & \text{Otherwise}.
\end{cases}.
\end{split}
\end{equation}
Given that  we use OFDMA in this setup, we have the following constraint \textcolor{black}{in which each subcarrier is assigned to at most one user}:
\begin{equation} \label{eqo2}
\text{\text{C1: }}\sum_{s \in \mathcal{S}}\sum_{u^s_j \in \mathcal{U}^s_j} x_{u^s_j}^{k} \le 1, \forall j\in\mathcal{J}, k\in\mathcal{K}.\nonumber
\end{equation}
The achievable rate of tactile user $u^s_j$ on subcarrier $k$ at BS $j$ can be calculated as
\begin{equation} \label{eqo3}
\begin{split}
 r_{u^s_j}= \sum_{k \in \mathcal{K}}  x_{u^s_j}^{k} \log_2(1+\gamma^{k}_{u^s_j}), \forall u^s_j \in \mathcal{U}^s_j, s \in \mathcal{S}, j \in \mathcal{J},
\end{split}
\end{equation}
where  
$\gamma^{k}_{u^s_j}=\frac {p_{u^s_j}^{k}h_{u^s_j}^{k}}{\sigma_{u^s_j}^{k}+I_{u^s_j}^{k}}$ is signal to interference plus noise ratio (SINR), in which $p_{u^s_j}^{k}$, $h_{u^s_j}^{k}$, and $\sigma_{u^s_j}^{k}$ represent the transmit power, channel power gain, and noise power from  BS $j$ to user $u^s_j$  on subcarrier $k$ for service $s$, respectively.
$ I_{u^s_j}^{k}$  is inter-cell interference which is equal to 
\begin{align}
& I_{u_j^s}^k = \sum\limits_{\scriptstyle m \in {\cal J},\hfill\atop
	\scriptstyle m \ne j\hfill} {\sum\limits_{v \in {\cal S}} {\sum\limits_{u_m^v \in {\cal U}_m^v} {x_{u_m^v}^k} } } p_{u_m^v}^k h_{u_m^v,j}^k, \nonumber\\& \forall u^s_j \in \mathcal{U}^s_j, s \in \mathcal{S}, j \in \mathcal{J}, k \in \mathcal{K}.\nonumber
\end{align}
where $h_{u_m^v,j}^k$ is the channel power gain between BS $j$ and user $u_m^v$ over subcarrier $k$.
\textcolor{black}{We introduce $\vartheta_{o_j}^{u^s_m} \in\{0,1\}$  which is set to $1$ if  teleoperator $o_j$ at BS $j$ is paired with tactile user  $u^s_m$ at BS $m$ with requested service $s$ i.e., tactile user  $u^s_m$ sends its data via UL transmission link to  teleoperator  $o_j$ over DL transmission link, which is paired.  Since we adopt FDD, the sets of UL and DL subcarriers are different.  As each tactile user has at most one paired teleoperator, we have
\begin{align}
\sum_{j \in \mathcal{J}}\sum_{o_j \in \mathcal{O}_j }\vartheta_{o_j}^{u^s_m} \le 1, \forall u^s_m \in \mathcal{U}^s_m, m \in \mathcal{J}, s\in \mathcal{S}. \nonumber
\end{align}}

For teleoperators, we consider a binary variable $\tau_{o_j}^{l}$ which is set to 1 if subcarrier $l$ is assigned to teleoperator $o_j$ at BS $j$ and otherwise is set to 0, i.e.,
\begin{equation} \nonumber
\begin{split}
&\tau_{o_j}^{l}=\\&\begin{cases}
1, &\text{If \text{subcarrier} $l$ is assigned to teleoperator $o_j$ at BS~$j$},\\
0, & \text{Otherwise}.
\end{cases}.
\end{split}
\end{equation}
Again, due to using OFDMA, we have the following constraint:
\begin{equation} \label{eqo2}
\text{\text{C2: }}\sum_{o_j \in \mathcal{O}_j} \tau_{o_j}^{l} \le 1, \forall j\in\mathcal{J}, l\in\mathcal{L}.\nonumber
\end{equation}
The achievable rate of teleoperator $o_j$  at BS $j$ can be calculated as
\begin{equation} \label{eqo3}
\begin{split}
&\hat r_{o_j}= \sum_{l \in \mathcal{L}}  \tau _{o_j}^{l}\log_2(1+\hat \gamma^{l}_{o_j}), \forall o_j \in \mathcal{O}_j, j \in \mathcal{J},
\end{split}
\end{equation}
where 
$\hat \gamma^{l}_{o_j}=\frac {\hat p_{o_j}^{l} \hat h_{o_j}^{l}}{\hat \sigma_{o_j}^{l}+\hat I_{o_j}^{l}}$ is SINR of teleoperator $o_j$ on subcarrier $l$, in which $\hat p_{o_j}^{l}$, $\hat h_{o_j}^{l}$, and $\hat \sigma_{o_j}^{l}$ represent the transmit power, channel power gain, and noise power from  teleoperator $o_j$ to BS $j$ over subcarrier $l$, respectively.
$\hat I_{o_j}^{l}$  is inter-cell interference which is equal to 
\begin{equation}
\hat I_{{o_j}}^l = \sum\limits_{\scriptstyle m \in {\cal J},\hfill\atop
	\scriptstyle m \ne j\hfill} {\sum\limits_{{o_m} \in {{\cal O}_m}} {\tau _{{t_m}}^l} } \hat p_{{t_m}}^l\hat h_{{o_j},m}^l, \forall {o_j}\in \mathcal{O}_j, j \in \mathcal{J},\nonumber
\end{equation}
where $\hat h_{{o_j},m}^l$ is the channel power gain between teleoperator $o_j$ and BS $j$. 

Due to the power limitation of each tactile user in UL transmission, we have
\begin{equation} \label{eqo72}
\text{\text{C3: }}\sum_{k \in \mathcal{K}}  x_{u^s_j}^{k}p_{u^s_j}^{k} \le P_{u^s_j} ^{\text{Max}}, \forall u^s_j\in\mathcal{U}^s_j, s \in \mathcal{S}, j \in \mathcal{J}.\nonumber
\end{equation}	
where $P_{u^s_j} ^{\text{Max}}$ is the maximum transmit power of tactile user ${u^s_j}$.
Moreover, due to the power limitation of each BS, we have
\begin{equation} \label{eqo73}
\text{\text{C4: }}\sum_{o_j \in \mathcal{O}_j} \sum_{l \in \mathcal{L}} \tau_{o_j}^{l} \hat p_{o_j}^{l} \le \hat P_j ^{\text{Max}}, \forall j\in\mathcal{J}.\nonumber
\end{equation}
where $\hat P_j ^{\text{Max}}$ is the maximum transmit power of BS $j$.	
\begin{figure*}[t]
	\centering
	\centerline{\includegraphics[width=1.1\textwidth]{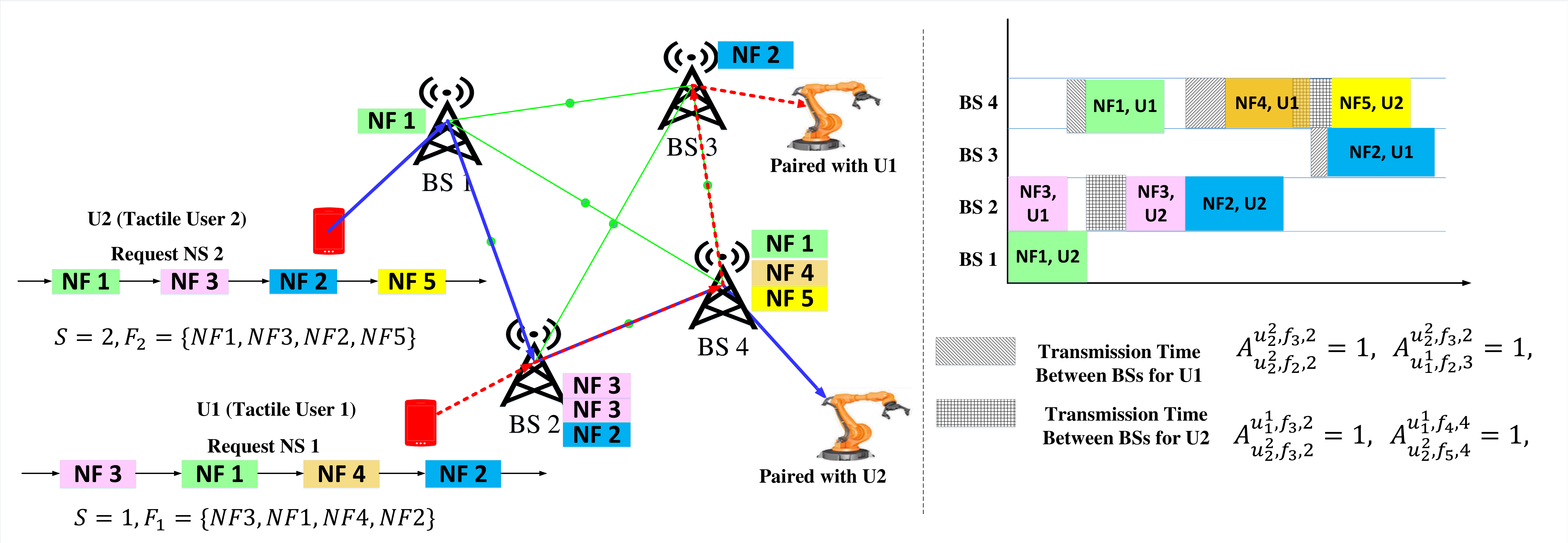}}
	\caption{ NFV Resource Allocation and Delays }
	\label{NFV_Pic}
\end{figure*}
\subsection{\textcolor{black}{NFV Resource Allocation}}
In this paper, we propose a general solution for placement and scheduling of VNFs.
An example of VNF placement and scheduling is shown in Fig. \ref{NFV_Pic}. To achieve this goal, we introduce a binary variable $A_{u^s_j,f_s,n_2}^{u^w_m,f_w,n_1}\in \{0,1\}$ which is set to $1$ if  NF $f_s$ of NS $s$ for tactile user $u^s_j$ is executed at BS $n_2$ after NF $f_w$ of NS $w$ for tactile user $u^w_m$ was executed at BS $n_1$ and is set to $0$ otherwise, i.e.,
\begin{equation} \nonumber
\begin{split}
& A_{u^s_j,f_s,n_2}^{u^w_m,f_w,n_1}\\&=\begin{cases}
1, &\text{If~NF~}f_s\text{~of~NS~}s\text{~for user~$u^s_j$~at~BS~}n_2\text{~is~executed}\\ &\text{~after~NF~}f_w\text{~of~NS~}w\text{~for~user~$u^w_m$~at~BS~}n_1,\\
0, & \text{otherwise}.
\end{cases}.
\end{split}
\end{equation}
It is assumed that each NF of each NS is performed at only one BS. Therefore, we have:
\begin{align} 
&\text{\text{C5: }} \sum_{j' \in \mathcal{J}} A_{u^s_{j},f_s,j'}^{u^w_m,f_w,n_1} \le 1, \\& \forall m,j \in \mathcal{J}, u^w_m, u^s_j \in \mathcal{U}^s_j, s,w \in \mathcal{S}, f_s \in \mathcal{F}_s, f_w \in \mathcal{F}_w. \nonumber
\end{align}

%
 \begin{figure*}[t]
	\centering
	\centerline{\includegraphics[width=0.8\textwidth]{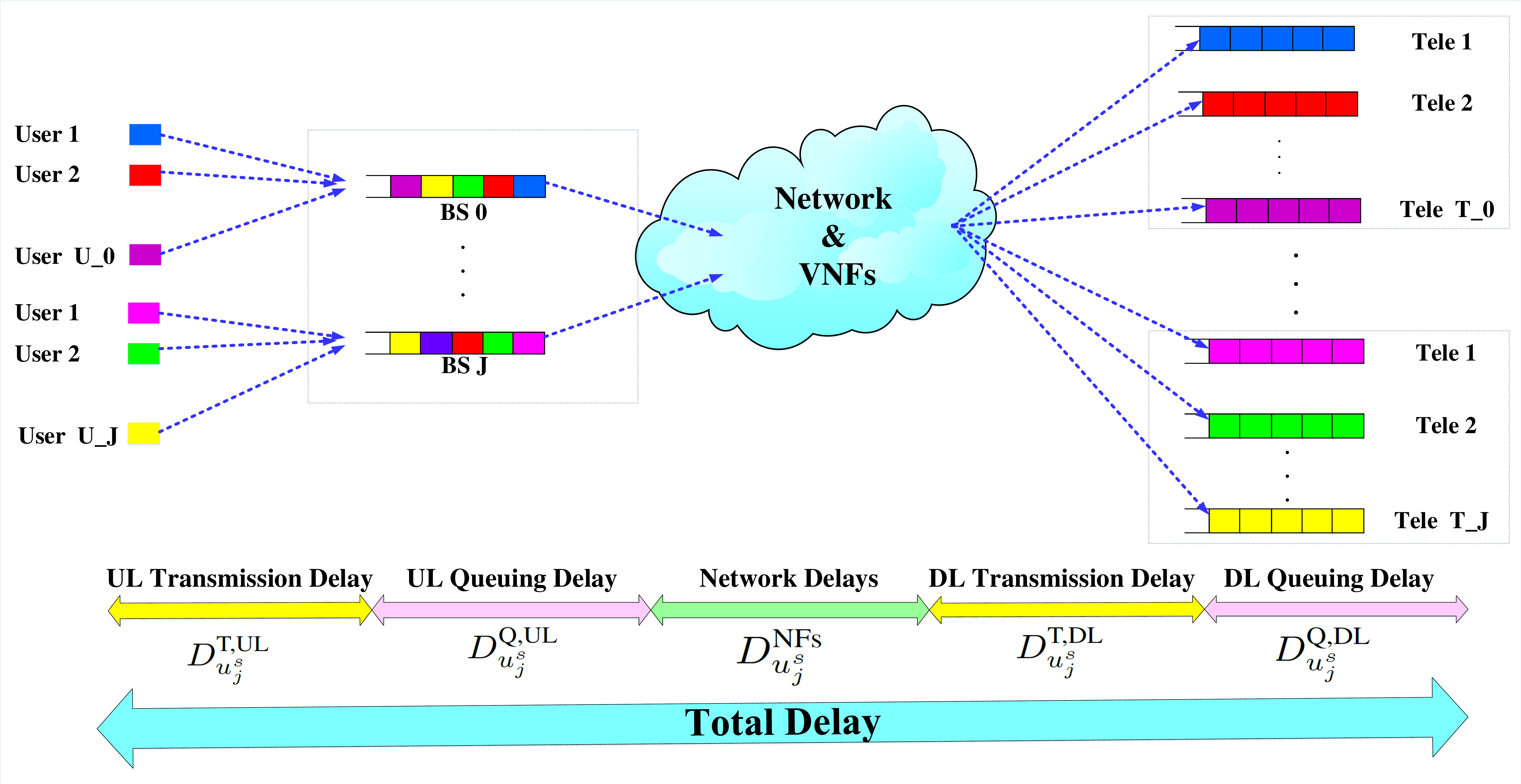}}
	\caption{ Delay Model }
	\label{pic1}
\end{figure*}
\subsection{Delay Model}

\textcolor{black}{The total delay for tactile user ${u^s_j}$ in our system model consists of five components as shown in Fig. \ref{pic1}, namely, 1) UL transmission delay in access ($D_{u^s_j}^{\text{T,UL}}$), 2) UL queuing delay ($D_{u^s_j}^{\text{Q,UL}}$), 3) delay resulting from executing  NFs ($D_{u^s_j}^{\text{NFs}}$), 4) DL transmission delay in access ($D_{u^s_j}^{\text{T,DL}}$), and 5) DL queuing delay ($D_{u^s_j}^{\text{Q,DL}}$). Due to the delay constraint for each NS, we have}
\begin{align}
&\text{C6: }D_{u^s_j}^{\text{T,UL}}+D_{u^s_j}^{\text{T,DL}}+D_{u^s_j}^{\text{NFs}}+D_{u^s_j}^{\text{Q,UL}}+D_{u^s_j}^{\text{Q,DL}}\le D_{u^s_j}^{{\rm{max}}}, \\&\forall u^s_j \in \mathcal{U}^s_j, j \in \mathcal{J}, s\in \mathcal{S}, \nonumber
\end{align}
where $D_{u^s_j}^{{\rm{max}}}$ is
maximum allowable E2E delay.
\subsubsection{\textbf{Transmission Delay}}
To calculate the UL and DL transmission delays, we have 
\begin{align} 
&\text{C7: } D_{u^s_j}^{\text{T,UL}} \ge \frac{C_{u^s_j}}{r_{u^s_j}}, u^s_j \in \mathcal{U}^s_j, s \in \mathcal{S}, j \in \mathcal{J}, \nonumber
\end{align}
\begin{align} 
\text{C8: } D_{u^s_j}^{\text{T,DL}} \ge  \frac{C_{u^s_j}}{\sum_{m\in \mathcal{J}}\sum_{o_m \in \mathcal{O}_m} \vartheta_{t_m}^{u^s_j}\hat r_{t_m}}, \nonumber \\   \forall  u^s_j \in \mathcal{U}^s_j, s \in \mathcal{S}, j \in \mathcal{J},\nonumber
\end{align}
where $C_{u}^{s}$ is the total transmitted bits of user $u^s_j$.
\subsubsection{\textbf{Delays of NS}}

To calculate the total delay of each NS for each user, we introduce a continuous variable $\omega_{u^s_j}^{f}$ which is set to end of the execution time of NF $f$ at BS $j$ for user $u^s_j$ with NS $s$.
In addition, we assume the link between BSs has limited capacity $\Psi_{n_1}^{n_2}, \forall n_1, n_2 \in \mathcal{J}$.
 Therefore, we have the following constraint
	\begin{align} \label{C99}
 		&\text{C9: } \small{ A_{u^s_j,f_s,n_2}^{u^b_m,f_b,n_1} \omega_{u^s_j}^{f} \ge A_{u^s_j,f_s,n_2}^{u^b_m,f_b,n_1} (\omega_{u^b_m}^{f_b} +\beta_{n_2}^{f_s}\frac{C_{u^s_j}}{\varOmega_{n_2}}+\mathds{1}_{n_2\neq n_1} \frac{C_{u^s_j}}{\Psi_{n_1}^{n_2}} ),}\nonumber\\& \forall u^s_j \in \mathcal{U}^s_j, u^b_m \in \mathcal{U}^b_m ,b,s\in \mathcal{S}, f_s \in \mathcal{F}_s, f_b \in \mathcal{F}_b,\nonumber\\& n_1,n_2,j,m \in \mathcal{J},\nonumber
 		\end{align}
\textcolor{black}{ where $\beta_{n_2}^{f_s}$ is the data processing coefficient of NF $f_s$ in BS $n_2$ and $\varOmega_j$ is the processing rate  of BS $j$ (bit/s).}
 \textcolor{black}{In this paper, we assume that the service requests of all users are received, then the scheduling and placement of VNFs are performed. Therefore, the delays associated to execution of NFs for each user is equal to the time when the last function  is executed.}
The total delay resulting from executing each NS is calculated as
\begin{align} 
	&\text{C10: }D_{u^s_j}^{\text{NFs}} \ge  \omega_{u^s_j}^{F_s}, \forall u^s_j \in \mathcal{U}^s_j.
\end{align} 
%

\subsubsection{\textbf{Queuing Delay}}

The queuing delay consists of two components: 1) UL queuing delay and 2) DL queuing delay, as shown in Fig. \ref{pic1}. Based on \cite{she2016ensuring}, the  aggregation of arrival bit rate of users can be modeled as a Poisson process. For a Poisson arrival process,  the effective bandwidth of tactile user $u^s_j$ is obtained as \cite{chang1995effective,she2016ensuring}
\begin{equation} \nonumber
B_{u^s_j}=\Lambda_{u^s_j} \frac{(e^{\theta_{u^s_j}}-1)}{\theta_{u^s_j}}, \forall _{u^s_j}\in\mathcal{U}^s_j, s \in \mathcal{S}, j \in \mathcal{J},
\end{equation}
where $\theta_{u^s_j}$ is the statistical QoS exponent of tactile user ${u^s_j}$. A larger  $\theta_{u^s_j}$ indicates a more stringent QoS and a smaller  $\theta_j$ implies a looser QoS requirement. $\Lambda_{u^s_j}$ is the number of bits arrived in time unit at BS $j$ queue from tactile user ${u^s_j}$ defined as 
$
{\Lambda _{u^s_j}} =r_{u^s_j}  ,~ \forall {u^s_j}\in\mathcal{U}^s_j$.
In the {TI} service, queuing delay violation probability should be ultra-low \cite{she2016ensuring}. Therefore, 
queuing delay violation probability for tactile user $u^s_j$ in UL transmission is obtained as follows
\begin{align} \label{deqo17}
&\epsilon_{u^s_j}^{\text{UL}}= \Pr\{D_{u^s_j}^{\text{UL}}>D_{u^s_j}^{\text{Q,UL}}\}=\eta_1 \exp(-\theta_{u^s_j} B_{u^s_j} D_{u^s_j}^{\text{Q,UL}})\le {\delta _1},\nonumber\\& \forall {u^s_j} \in\mathcal{U}^s_j , s \in \mathcal{S}, j \in \mathcal{J},
\end{align}
 where $D_{u^s_j}^{\text{UL}}$ is the ${u^s_j}^{\text{th}}$ tactile user queuing delay in UL, $D_{u^s_j}^{\text{Q,UL}}$ is the maximum queuing delay in UL,  $\eta_1$ is the  non-empty buffer probability in UL, and $\delta _1$ is the maximum queuing delay violation probability in UL. Equation \eqref{deqo17} can be simplified to
\begin{equation}\nonumber
\begin{split}
&\exp ( - {\theta_{u^s_j}}B^j{D_{u^s_j}^{\text{Q,UL}}}) = \exp ( - {\theta_{u^s_j}}{\Lambda_{u^s_j}}\frac{{({e^{{\theta_{u^s_j}}}} - 1)}}{{{\theta_{u^s_j}}}}{D_{u^s_j}^{\text{Q,UL}}})=\nonumber\\&  \exp (- {\Lambda_{u^s_j}}({e^{{\theta_{u^s_j}}}} - 1){D_{u^s_j}^{\text{Q,UL}}}) \le {\delta _1}.
\end{split}
\end{equation}
Therefore, we have\footnote{In this work, we have two buffers for each user: the first one is for the uplink (UL) transmission and the second one is for the downlink (DL) transmission. For the queue in the UL transmission, we model the delay based on the effective bandwidth which leads to a constraint on the UL rate and for the second one we model the delay based on the effective capacity which leads to a constraint on DL rate. This model is similar to \cite{aijaz2016towards,4543084,1210731}.}	
\begin{equation}
\begin{split}
{\text{C11: }} r_{u^s_j}  \ge \frac{{\ln ({1/\delta _1})}}{{( {e^{{\theta_{u^s_j}}}}-1){D_{u^s_j}^{\text{Q,UL}}}}}, \forall _{u^s_j} \in \mathcal{U}^s_j, s \in \mathcal{S}, j \in \mathcal{J}.\nonumber
\end{split}
\end{equation}
	
Similarly, in DL transmission, we have
\begin{align} \label{deqo18}
&\epsilon_{u^s_j}^{\text{DL}}= \Pr\{ D_{u^s_j}^{\text{DL}}> D_{u^s_j}^{\text{Q,DL}}\}=\eta_2 \exp(-\tilde \theta_{u^s_j} \tilde B_{u^s_j}  D_{u^s_j}^{\text{Q,DL}}),\nonumber\\&  \forall _{u^s_j} \in \mathcal{U}^s_j, s \in \mathcal{S}, j \in \mathcal{J},
\end{align}
where $ D_{u^s_j}^{\text{DL}}$ is the ${u^s_j}^{\text{th}}$ tactile user at BS $j$ queuing delay in DL, $ D_{u^s_j}^{\text{Q,DL}}$ is the maximum queuing delay in DL, $\eta_2$ is the  non-empty buffer probability in DL, and $\delta _2$ is the maximum queuing delay violation probability in DL. Therefore, Equation \eqref{deqo18} can be simplified to	
\begin{equation}\nonumber
\begin{split}
&	\exp ( - {\tilde \theta_{u^s_j}}\tilde B_{u^s_j}{ D_{u^s_j}^{\text{Q,DL}}}) = \exp ( - {\tilde \theta_{u^s_j}}{\tilde \Lambda _{u^s_j}}\frac{{({e^{{\tilde \theta _{u^s_j}}}} - 1)}}{{{\tilde \theta_{u^s_j}}}}{ D_{u^s_j}^{\text{Q,DL}}})=\nonumber\\&  \exp (- {\tilde \Lambda _{u^s_j}}({e^{{\tilde \theta_{u^s_j}}}} - 1){ D_{u^s_j}^{\text{Q,DL}}}) \le {\delta _2}.
\end{split}
\end{equation}
Therefore, we have	
\begin{align}
{\text{C12: }} &\sum_{m \in \mathcal{J}}\sum_{o_m \in \mathcal{O}_m } \vartheta_{o_m}^{u^s_j}\hat r_{o_m}  \ge \frac{{\ln ({1/\delta _2})}}{{( {e^{{\tilde \theta_{u^s_j}}}}-1){ D_{u^s_j}^{\text{Q,DL}}}}},\nonumber\\&  \forall u^s_j \in \mathcal{U}^s_j,  s \in \mathcal{S}, j \in \mathcal{J}.\nonumber
\end{align}
\begin{algorithm}[t]
	\caption{Three-Step Iterative Algorithm}	
	{\textbf{Step 1: Initialization}}
	\begin{itemize}
		\item [] $\mathcal{J} = \{1,...,J\}$,$\mathcal{K}_1 = \{1,...,K\}$, $\mathcal{L} = \{1,...,L\}$, $\mathcal{U}_j^s = \{1,...,U_j^s\}$, $\mathcal{S} = \{1,...,S\}$, $\epsilon_{\text{TH}}=10^{-4}$, $Z_{\text{TH}}=100$
		and $z=0$, Initial value $\dot p^{(z)}=\dot p^0$,  $\ddot p^{(z)}=\ddot p^0$.
	\end{itemize} 
	{\textbf{Step 2: Solving the optimization problem \eqref{eqoa1}}}
	\begin{itemize}
		\item [] {\textbf{phase 1: Subcarrier Allocation:}} Allocate subcarrier by minimizing the total cost  and satisfying the constraints of problem \eqref{eqoa1}
		\item [] {\textbf{phase 2: Power Allocation:}}	Allocate power to each user according to problem (\ref{eqoa1}) and subcarriers allocated in phase 1.
		\item [] {\textbf{phase 3: NFV Resource Allocation:}}Embedding and Scheduling done according to problem (\ref{eqoa1}).
		\item [] {\textbf{phase 4: Delay Adjustment:}} Adjust delay  according to problem (\ref{eqoa1}).
	\end{itemize}
 {\textbf{Step 3: Iteration}}
 \begin{itemize}
 	\item [] $z=z+1$, Repeat Step 2, 3, 4 and 5 until $||\boldsymbol{\dot P}^{(z)}-\boldsymbol{\dot P}^{(z-1)}||\le\epsilon_{\text{TH}}$ and $||\boldsymbol{\ddot P}^{(z)}-\boldsymbol{\ddot P}^{(z-1)}||\le\epsilon_{\text{TH}}$ or  $Z_{\text{TH}}<z$.
 \end{itemize}
	\label{ALG1}
\end{algorithm}
\section{Optimization Problem Formulation and Solution}\label{optimizationsolve} 
 \textcolor{black}{In the proposed system model, we define the cost function $C$ as the total amount of physical and virtual network resource occupied for resource allocation, i.e.,}
	\begin{align}
&	C=\varrho_1 \sum_{j \in \mathcal{J}}\bigg(\sum_{s \in \mathcal{S}} \sum_{{u^s_j} \in \mathcal{U}^s_j}\sum_{k \in \mathcal{K}}x_{u^s_j}^{k} p_{u^s_j}^{k}+ \sum_{o_j \in \mathcal{O}_j} \sum_{l \in \mathcal{L}} \tau_{o_j}^{l} \hat p_{o_j}^{l}\bigg)+\nonumber\\&\varrho_2 \bigg( \sum_{j \in \mathcal{J}} \sum_{s \in \mathcal{S}} \sum_{{u^s_j} \in \mathcal{U}^s_j} \sum_{n_2 \in \mathcal{J}}  \max_{u^b_m,f_b,n_1}\{A_{u^s_j,f_s,n_2}^{u^b_m,f_b,n_1}\} \beta_{n_2}^{f_s}\frac{C_{u^s_j}}{\varOmega_{n_2}} \bigg) \nonumber,
	\end{align}
	where $\varrho_1$ and $\varrho_2$ are constant values for scaling the transmit power and execution time of NFs 
	and have units such that the overall costs can be expressed in terms of monetary value.	
		In the simulation setup, we set  $\varrho_1$   to 1\$/watts and  $\varrho_2$ to 1\$/milliseconds. It is worth noting that the exact choices of the weights  $\varrho_1$ and  $\varrho_2$ depend on the business model of the network operator. 
 The aim is to minimize the cost function by considering an E2E delay constraint and limitation of radio resources in the network. Based on the mentioned constraints C1-C12, the optimization problem can be written as
 \vspace{-0.5em}
	\begin{align}\label{eqoa1}
	&\mathop {\min }\limits_{\scriptstyle{\bf{ P}},{\bf{T,X}},\hfill\atop
		\scriptstyle{\bf{W}},{\bf{A}},{\bf{D}}\hfill} C \nonumber
	\\\text{s.t.:}&\text{~(C1)-(C12)}.
	\end{align}
	The optimization variables in \eqref{eqoa1} are subcarrier allocation, power allocation, NFV allocation, time allocation, and delay adjustment for different users in access  as well as in both UL and DL where $\boldsymbol{P}$,  $\boldsymbol{X}$,$\boldsymbol{T}$, $\boldsymbol{W}$, $\boldsymbol{A}$, and $\boldsymbol{D}$ are the transmit power, the tactile users subcarrier allocation, teleoperators subcarrier  allocation, end time vector of VNF executions, VNF scheduling vector, and delay adjustment, respectively.
	In problem \eqref{eqoa1}, the rate function is a non-convex function, which leads to the non-convexity of the problem. In addition, this problem contains both discrete and continuous variables, which makes the problem more challenging. 

The optimization problem \eqref{eqoa1} is mixed-integer non-convex programming and hard to solve. In order to facilitate the solution of this problem, we define new variables as $\dot p_{u^s_j}^{k}=x_{u^s_j}^{k} p_{u^s_j}^{k}$ and $\ddot p_{o_j}^{l}=\tau_{o_j}^{l} \hat p_{o_j}^{l} $. Therefore,  \eqref{eqoa1} is turned into the following problem
 \begin{align}\label{eqoa_Total}
 &\mathop {\min }\limits_{\scriptstyle{\bf{\dot P}},{\bf{\ddot P}},{\bf{T},\bf{X}},\hfill\atop
 	\scriptstyle{\bf{D}},{\bf{W}},{\bf{A}}\hfill}\varrho_1\sum_{j \in \mathcal{J}} \bigg(\sum_{s \in \mathcal{S}} \sum_{{u^s_j} \in \mathcal{U}^s_j}\sum_{k \in \mathcal{K}} \dot p_{u^s_j}^{k}+ \sum_{o_j \in \mathcal{O}_j} \sum_{l \in \mathcal{L}}\ddot p_{o_j}^{l}\bigg)\nonumber\\&+\varrho_2 \bigg( \sum_{j \in \mathcal{J}} \sum_{s \in \mathcal{S}} \sum_{{u^s_j} \in \mathcal{U}^s_j} \sum_{n_2 \in \mathcal{J}}  \max_{u^b_m,v,n_1}\{A_{u^s_j,f_s,n_2}^{u^b_m,f_b,n_1}\} \beta_{n_2}^{f_s}\frac{C_{u^s_j}}{\varOmega_{n_2}} \bigg) \nonumber
 \\ \text{s.t.:}& \text{~(\~{C1})-(\~{C12})}.
 \end{align}
  To solve problem \eqref{eqoa_Total}, we exploit alternative search method (ASM) which is described in Algorithm \ref{ALG1} \cite{bertsekas1997nonlinear, wang2011iterative}. Thus, the approach is to divide the problem into 4 subproblems: 1) Subcarrier allocation subproblem, 2) Power allocation subproblem, 3) NFV-RA subproblem, 4) Delay adjustment subproblem. These subproblems are solved iteratively, in which $z$ represents the iteration number. Please note that although the optimization process is performed thorough solving 4 subproblems, it is performed sequentially within an iterative framework such that the solution of each subproblem directly affects that of the next. As such, the resources are still allocated jointly in contrast to the separate approach (SA) case.

  First, the initial values for power allocation, subcarrier allocations, NFV-RA, and delay adjustment should be obtained. Next, in each iteration, we obtain the optimal solution for the subcarrier allocation subproblem  with $\dot p^{(z)}=\dot p^0$ and  $\ddot p^{(z)}=\ddot p^0$. Then an optimal solution for the power allocation subproblem is obtained with  $ X^{(z)}=x^0$ and  $T^{(z)}=\tau^0$ and the solution is fed as the initial value tonb the next subproblem. Later in the NFV-RA subproblem, we obtain a solution of the VNF placement and scheduling to satisfy problem \eqref{eqoa1} constraints. After that, we solve the delay adjustment subproblem and delay is adjusted by solving an integer linear programming (ILP) problem.    
  The iteration stops when the iteration number exceeds a predetermined value $Z_{\text{TH}}$ or $||\boldsymbol{\dot P}^{(z)}-\boldsymbol{\dot P}^{(z-1)}||\le\epsilon_{\text{TH}}$ and $||\boldsymbol{\ddot P}^{(z)}-\boldsymbol{\ddot P}^{(z-1)}||\le\epsilon_{\text{TH}}$, where $||.||$ represents  the vector Euclidean norm. 
  The output of the last iteration is the suboptimal solution of Problem \eqref{eqoa1}.  \textcolor{black}{It should be noted that there is no local or global optimality guarantee for ASM.}

 \subsection{Subcarrier Allocation Subproblem}
The subcarrier allocation subproblem is as follows:
\begin{align}\label{eqoS1}
&\text{find~} ({\boldsymbol{T},\boldsymbol{X}}) 
\\ \text{s.t.}:&\text{(C1)-(C2)}. \nonumber
\end{align}
\textcolor{black}{In this subproblem, if the subcarrier with the highest channel gain is allocated  to each user, the total transmit power is reduced and the total rate is increased which affects the delay (by increasing the total rate, the delay is reduced). Therefore, to avoid this issue,} we propose a heuristic approach summarized in Algorithm \ref{ALGS}.

 \begin{algorithm}[t]
 	\caption{Heuristic Algorithm  for the Subcarrier Allocation Subproblem}	
 	\label{Gale-Shapley algorithm}
 		\textbf{Step 1: \textit{Initialization}}
 		\begin{itemize}
 			\item Ascendingly sort  $u^s_j$ in BS $j$  according to the required  delay  $D_{u^s_j}^{{\rm{max}}}$ in vector $\hat U$  
 			\item Ascendingly sort  $o_j$ in BS $j$ according to the required delay  $\sum_{j \in \mathcal{J}}\sum_{o_j \in \mathcal{O}_j }\vartheta_{o_j}^{u^s_m} D_{u^s_j}^{{\rm{max}}}$ in vector $\hat O$  
 			\item Descendingly sort subcarriers for each $u^s_j$ according to $\dot p_{u^s_j}^{k}h_{u^s_j}^{k}$  in vector $\hat K$ 
 			\item Descendingly sort subcarriers for each $o_j$ according to $\ddot p_{o_j}^{l}\hat h_{o_j}^{l}$  in vector $\hat L$
 		\end{itemize}
 			\textbf{Step 2: \textit{Subcarrier Allocation}}  
 			\begin{enumerate}
 				\item For each user $ u^s_j \in  {\hat U}$ based on priority, find the best subcarrier.
 				\begin{enumerate}
 					\item Each subcarrier is assigned at most to one user
 				\end{enumerate}
 				\item For each teleoperators $ o_j \in  {\hat O}$ based on priority,  find the best subcarrier.
 				\begin{enumerate}
 					\item Each subcarrier is assigned at most to one teleoperator
 				\end{enumerate}
 				\item Check constraint C6-C12 of problem \eqref{eqoa_Total} for all users
 				\begin{enumerate}
 					\item If it is satisfied, go to Step 3
 					\item Else go back to Step 2 and execute the following item:
 					\begin{enumerate}
 						\item Change the order of user for which the delay constraint is not satisfied  with one of previous users whose delay requirement is satisfied in a way that the total delay is considerably less than the maximum allowable amount according to the constraint C6.
 					\end{enumerate}
 				\end{enumerate}
 			\end{enumerate}
 				 	\textbf{Step 3: \textit{Output:}} ${\boldsymbol{T},\boldsymbol{X}}$ 
 	\label{ALGS}
 \end{algorithm}
 \subsection{Power Allocation Subproblem}
The power allocation subproblem is as follows:
\begin{align} \label{eqoP}
&\min_{\boldsymbol{\dot P, \ddot P}}\varrho_1 \sum_{j \in \mathcal{J}}(\sum_{s \in \mathcal{S}} \sum_{{u^s_j} \in \mathcal{U}^s_j}\sum_{k \in \mathcal{K}} \dot p_{u^s_j}^{k}+ \sum_{o_j \in \mathcal{O}_j} \sum_{l \in \mathcal{L}}\ddot p_{o_j}^{l})
\\ \text{s.t.}:&\text{\~{C3}-\~{C4}},\text{\~{C7}-\~{C8}}, \text{\~{C11}-\~{C12}}. \nonumber
\end{align}
\textcolor{black}{ In \eqref{eqoP}, the constraints containing the rate function are non-convex. We convert them to convex constraints by applying the difference of two concave functions (DC) approximation which is based on the successive convex approximation (SCA) approach \cite{6678362}}. Hence, we write the rate function as follows:
\begin{align}
\begin{split}
	& r_{u^s_j}=\tilde f(\boldsymbol{\dot P})\\&= \sum_{k \in \mathcal{K}}  \log_2(1+\frac {\dot p_{u^s_j}^{k}h_{u^s_j}^{k}}{\sigma_{u^s_j}^{k}+ \sum\limits_{\scriptstyle m \in {\cal J},\hfill\atop
			\scriptstyle m \ne j\hfill} {\sum\limits_{v \in {\cal S}} {\sum\limits_{u_m^v \in {\cal U}_m^v}  } }\dot p_{u_m^v}^k h_{u_m^v,j}^k})\\&=\sum_{k \in \mathcal{K}} \log_2(\sigma_{u^s_j}^{k}+ \sum\limits_{\scriptstyle m \in {\cal J},\hfill\atop
		\scriptstyle m \ne j\hfill} {\sum\limits_{v \in {\cal S}} {\sum\limits_{u_m^v \in {\cal U}_m^v}  } } \dot p_{u_m^v}^k h_{u_m^v,j}^k+ {\dot p_{u^s_j}^{k}h_{u^s_j}^{k}})\\&-  \log_2(\sigma_{u^s_j}^{k}+ \sum\limits_{\scriptstyle m \in {\cal J},\hfill\atop
		\scriptstyle m \ne j\hfill} {\sum\limits_{v \in {\cal S}} {\sum\limits_{u_m^v \in {\cal U}_m^v}  } }\dot p_{u_m^v}^k h_{u_m^v,j}^k)\\&=f(\boldsymbol{\dot P})-g(\boldsymbol{\dot P}),\nonumber
\end{split}
\end{align}
where 
\begin{align}
& f(\boldsymbol{\dot P})=\\&\sum_{k \in \mathcal{K}} \log_2(\sigma_{u^s_j}^{k}+ \sum\limits_{\scriptstyle m \in {\cal J},\hfill\atop
	\scriptstyle m \ne j\hfill} {\sum\limits_{v \in {\cal S}} {\sum\limits_{u_m^v \in {\cal U}_m^v}  } } \dot p_{u_m^v}^k h_{u_m^v,j}^k+ {\dot p_{u^s_j}^{k}h_{u^s_j}^{k}}),\nonumber
\end{align}
\begin{equation}
\begin{split}
g(\boldsymbol{\dot P})=\log_2(\sigma_{u^s_j}^{k}+ \sum\limits_{\scriptstyle m \in {\cal J},\hfill\atop
	\scriptstyle m \ne j\hfill} {\sum\limits_{v \in {\cal S}} {\sum\limits_{u_m^v \in {\cal U}_m^v}  } }\dot p_{u_m^v}^k h_{u_m^v,j}^k).
\end{split}
\end{equation}

For each  user $u^s_j$,  we employ the following linear approximation based on first order Taylor series in the point ${\boldsymbol{\dot P}}^z$ as follows
\begin{equation}
\begin{split}
g(\boldsymbol{\dot P}^z)=g(\boldsymbol{\dot P}^{z-1})+\nabla g(\boldsymbol{\dot P}^{z-1})({{\boldsymbol{\dot P}}^z}-{\boldsymbol{\dot P}}^{z-1}),
\end{split}
\end{equation}
where $z$ indicates the iteration numbers and $\nabla g(\boldsymbol{\dot P}^{z-1})$ is obtained as follows:
\begin{equation}
\begin{split}
\nabla g(\boldsymbol{\dot P}^{z-1})= \left\{ {\begin{array}{*{20}{c}}
	0,&{{\rm{if~~~}}m = j},\\
	\frac{h_{u_j^s}^{k}}{{\sigma _{u_j^s}^{k} + I_{u_j^s}^{k} }},&{{\rm{if~}}m \ne j}.
	\end{array}} \right.
\end{split}
\end{equation}
 \subsection{NFV Resource Allocation Subproblem}
The NFV-RA subproblem is as follows:
	\begin{equation}
	\begin{split}\label{NFVP-1}
	&\min_{{\bf{W}},{\bf{A}}} \varrho_2 \bigg( \sum_{j \in \mathcal{J}} \sum_{s \in \mathcal{S}} \sum_{{u^s_j} \in \mathcal{U}^s_j} \sum_{n_2 \in \mathcal{J}}  \max_{u^b_m,v,n_1}\{A_{u^s_j,f_s,n_2}^{u^b_m,f_b,n_1}\} \beta_{n_2}^{f_s}\frac{C_{u^s_j}}{\varOmega_{n_2}} \bigg) \\
	\text{s.t.:}& \text{\text{ C5-C6, C9-C10. }}	
	\end{split}
	\end{equation}

 \textcolor{black}{Similar to the subcarrier allocation subproblem, we use a heuristic method described in Algorithm \ref{ALGNFV}.  It is worth noting that in JA approach, the decision (change the order) is made based on constraint C6 and in SA approach, the decision is made based on constraint C10.}
 \begin{algorithm}[t]
 	\caption{Heuristic Algorithm for NFV Resource Allocation}	
 	\label{Gale-Shapley algorithms}
  		\textbf{Step 1: \textit{Initialization}}
  		\begin{itemize}
 	\item Create matrix  $ \Psi_{n_1}^{n_2}, \forall n_1 ,n_2 \in \mathcal{J}, n_1\neq n_2$ for capacity between the links.  
 		\item Create vector $\varOmega_{n},  \forall n \in \mathcal{J}$ for processing capacity of each BS. 
 		\item For each NF of user $u^s_j$ calculate the processing time in each node $\tau(u^s_j,f_s,n)=\beta_{n}^{f_s}\frac{C_{u^s_j}}{\varOmega_{n}}$.
 		\item Ascendingly sort  $\tau(u^s_j,f_s,n)$ for each NF of user $u^s_j$.
 		\item For each user $u^s_j$ calculate transmission time from BS $j$ to BS $j'\neq j$ as $\zeta(u^s_j,j')=\frac{C_{u^s_j}}{ \Psi_{j}^{j'}}$.
 		\item Ascendingly sort $u^s_j$ according to the required delay  $D_{u^s_j}^{{\rm{max}}}$  in vector $\tilde U$. 
 	\end{itemize}
 		\textbf{Step 2: \textit{Placement Procedure}}
 		\begin{enumerate}
 			\item For each user $ u^s_j \in  {\tilde U}$ based on $\zeta(u^s_j,j')$ and  $\tau(u^s_j,f_s,n)$,  find the best base station for processing all of NFs.
 			\begin{enumerate}
 				\item Update $\omega_{u^s_j}^{f_s}$ and $A_{u^s_j,f_s,n_2}^{e^w_m,f_w,n_1}$ for all $u^{s'}_{j'}\le u^s_j, u^{s'}_{j'}, u^s_j \in \tilde U$
 				\item Calculate $D_{u^s_j}^{{\rm{max}}}-\omega_{u^s_j}^{F_s}$.
 			\end{enumerate} 
 			\item Check constraint C10 of problem \eqref{eqoa_Total} for all users:
 			\begin{enumerate}
 				\item  If it is satisfied, go to Step 3,
 				\item Else go back to Step 2 and execute the following item:
 				\begin{enumerate}
 					\item Change the order of user for which the delay constraint is not satisfied with one of the previous users whose delay requirement is satisfied in a way the total delay is considerably less than the maximum allowable desired amount  (constraint C6 for JA and C10 for SA).
 				\end{enumerate}
 			\end{enumerate}
 		\end{enumerate}
 	\textbf{Step 3: \textit{Output:}} ${\bf{W}},{\bf{A}}$
 	\label{ALGNFV}
 \end{algorithm}
 \subsection{Delay Adjustment Subproblem}
The delay adjustment subproblem is as follows:
	\begin{align}\label{NFVP-2}
&\text{find~}({\boldsymbol{D}}) 
\\ \nonumber \text{s.t.:}& \text{\text{ C6-C8, C10-C12}}.
	\end{align}
The delay adjustment subproblem is a LP problem and can be solved by any existing optimization toolbox such as CVX in MATLAB.

Here, we provide a discussion on the convergence of the proposed solution. For the solution to converge, it is necessary to examine the convergence of  Algorithm \ref{ALG1} and the convergence of each subproblem solution. In  Algorithm \ref{ALG1}, at first, the subcarrier subproblem is  solved. In this subproblem, to each user the subcarrier with maximum channel gain is allocated according to the initial power value. The ordering of subcarrier allocation is based on the priority of the requested service by the user in terms of the E2E delay requirement. If the delay constraints are not met for some users, the priority of those users will change. It should be noted that this subproblem does not directly affect the objective function as far as convergence of main problem is concerned. Next, the power allocation subproblem is solved using the SCA method which is a convergent algorithm  \cite{wang2011iterative,mokdad2016radio}. Therefore, in each iteration, we have $f({\bf{\dot P}}^{z-1},{\bf{\ddot P}}^{z-1}) \ge f({\bf{\dot P}}^{z},{\bf{\ddot P}}^z)$. The next subproblem is the NFV-RA subproblem. The VNFs  placement and scheduling are performed based on NS priority in terms of E2E delay. It should be noted that NFV-RA subproblem does not directly affect the objective function as far as convergence of the main problem is concerned.  The last subproblem is the delay adjustment subproblem whose convergence is intuitive. Therefore, in each iteration, a feasible value for power is obtained, which is less than or equal to the feasible value in the previous iteration and the overall algorithm converges \textcolor{black}{in the achieved performance (not necessarily in the optimization variables)} \cite{rezvani2019fairness,wang2011iterative,mokdad2016radio}.

\section{Computational complexity} \label{complexity}
In this section, the computational complexity of the proposed RA algorithm (Algorithm. \ref{ALG1}) is calculated. The overall RA
algorithm consists of four stages: (1)  the subcarrier allocation subproblem \eqref{eqoS1}, (2) the power allocation subproblem \eqref{eqoP}, (3) the NFV-RA subproblem \eqref{NFVP-1}, and (4) the delay adjustment subproblem which are jointly considered through an iterative approach.  The computational complexity of each subproblem is shown in Table \ref{table-C} where $z_1$ and $z_2$ are the number of iterations in corresponding subproblem \cite{mokdad2016radio,moltafet2018optimal}. For power allocation subproblem and delay adjustment subproblem which are solved based on interior point method (IPM) by CVX in MATLAB, the computational complexity can be obtained as follows:
\begin{equation}
\dfrac{\log(Q_\text{Constr})/q^0 \rho}{\log \zeta},
\end{equation}
where $Q_\text{Constr}$ is equal to the total number of constraints,   $0\le \rho\le1$ is the stopping criterion of IPM, $\zeta$ is used to update accuracy of the IPM, and $q^0$ is initial point for approximating the accuracy of IPM.
\begin{table}[]
			\centering
			\caption{Computational Complexity}
			\label{table-C}
	\begin{tabular}{|c|c|}
		\hline
		\rowcolor[HTML]{96FFFB} 
		\textbf{Subproblems}             & \textbf{Computational Complexity}                                                                                                                                            \\ \hline
		Subcarrier Allocation subproblem & \begin{tabular}[c]{@{}c@{}}$\mathcal{O}\big( z_1  \times(U_ \text{Tot} K+4U_ \text{Tot}+U_ \text{Tot}^2$\\ $+O_ \text{Tot}L+4O_ \text{Tot}+O_ \text{Tot}^2)\big)$\end{tabular} \\ \hline
		Power Allocation subproblem      & $\mathcal{O}\left(\dfrac{\log(5U_ \text{Tot}+J)/q^0 \rho}{\log \zeta}\right)$                                                                                                \\ \hline
		NFV Allocation subproblem        & $\mathcal{O}\big(z_2\times(3U_ \text{Tot} F+U_ \text{Tot}^2 F^2 J^2+U_ \text{Tot})\big) $                                                                                    \\ \hline
		Delay adjustment subproblem      & $\mathcal{O}\left(\dfrac{\log(6U_ \text{Tot})/\tilde q^0 \tilde \rho}{\log\tilde  \zeta}\right)$                                                                             \\ \hline
	\end{tabular}
\end{table}
\section{Simulation Results} \label{simulationresults}

	\begin{table}[t]
		\centering
		\caption{Considered parameters in numerical results}
		\label{table-5}
		\begin{tabular}{|c|c|}
			\hline
			\rowcolor[HTML]{96FFFB} 
			\textbf{Parameter}                          & \textbf{Value of each parameter}      \\ \hline
			Coverage area                               & $10$$\sim$$\text{Km}^2$               \\ \hline
			Path loss exponent                          & $3$                                   \\ \hline
			QoS exponent                                & $11$                                  \\ \hline
			PSD  of the received AWGN noise             & $-174$ dBm/Hz                         \\ \hline
			Bandwidth of UL                             & $5$~MHz                          \\ \hline
			Bandwidth of DL                             & $5$~MHz                          \\ \hline
			Number of UL subcarriers                    & $8$                                   \\ \hline
			Number of DL subcarriers                    & $16$                                  \\ \hline
			Maximum transmit power of the MBS           & $46$~dBm                         \\ \hline
			Maximum transmit power of the SBS           & $43$~dBm                         \\ \hline
			Maximum transmit power of each tactile user & $23$~dBm                         \\ \hline
			$\varrho_1$                                 & $1\$/\text{watts}$ \\ \hline
			$\varrho_2$                                 & $1\$/\text{ms}$    \\ \hline
		\end{tabular}
	\end{table}
	\subsection{Simulation Setup}
In this section, the simulation results are presented to evaluate the performance of the proposed system model. In our simulations, the SBSs are located at an equal distance from MBS. The coverage area is considered $10$~$\text{Km}^2$. Moreover, we consider a  Rayleigh fading wireless channel in which the channel power gain of subcarriers are independent. Channel power gains for the radio access links are set as $h_{u^s_j}^{k}=\Omega_{u^s_j}^{k}{d_{u^s_j}}^\alpha$ where $d_{u^s_j}$ is the distance between user $u^s_j$ and BS $j$, $\Omega_{u^s_j}^{k}$ is a random variable with Rayleigh distribution, and $\alpha=3$ is the path-loss exponent. The power spectral density (PSD) of the additive white Gaussian noise (AWGN) is set to be $-174$~dBm/Hz. The system model parameters are summarized in Table \ref{table-5} 	\cite{ngo2014joint}.
	
	 \begin{figure}[t]
	 	\centering
	 	\centerline{\includegraphics[width=0.47\textwidth]{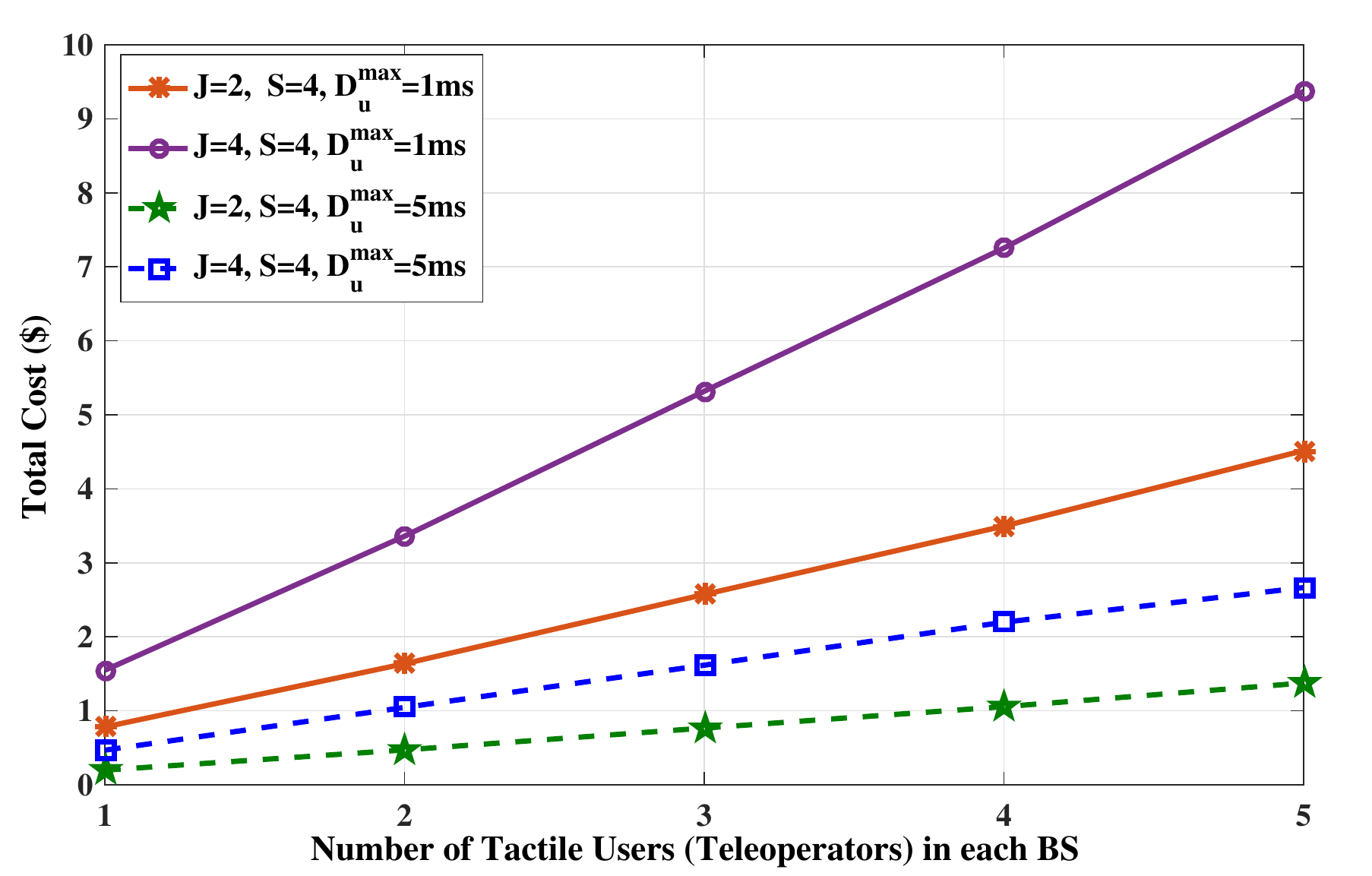}}
	 	\caption{ Total cost versus the number of users in each BS }
	 	\label{Fig1}
	 \end{figure}
	 \subsection{Performance Analysis}	
At first, we study the effects of network parameters on the performance of the proposed system model. In Fig. \ref{Fig1}, the total cost  is shown versus the number of tactile users in each BS for different NS requirements. As can be seen, by increasing the number of tactile users, the total cost increases. Moreover, by increasing the number of BSs, the number of served tactile users and subsequently the total cost increases.  The most important point is that as the NS E2E delay is decreased, the amount of total transmit power is increased and consequently the total cost is increased. Now if all tactile users request only one type of service with specific E2E delay requirement (e.g., $10$~ms or $1$~ms), for services with $1$~ms E2E delay requirement in the fixed-point, e.g., $J=4$ and $U=5$,  the total cost is about $9.3$ while in the setup with $5$~ms E2E delay requirement, the total cost is about $2.7$. Therefore, by decreasing the E2E delay requirement, the total transmit power increases drastically, and hence the total cost increases.

\begin{figure}[t]
	  	\centering
	  	\centerline{\includegraphics[width=0.5\textwidth]{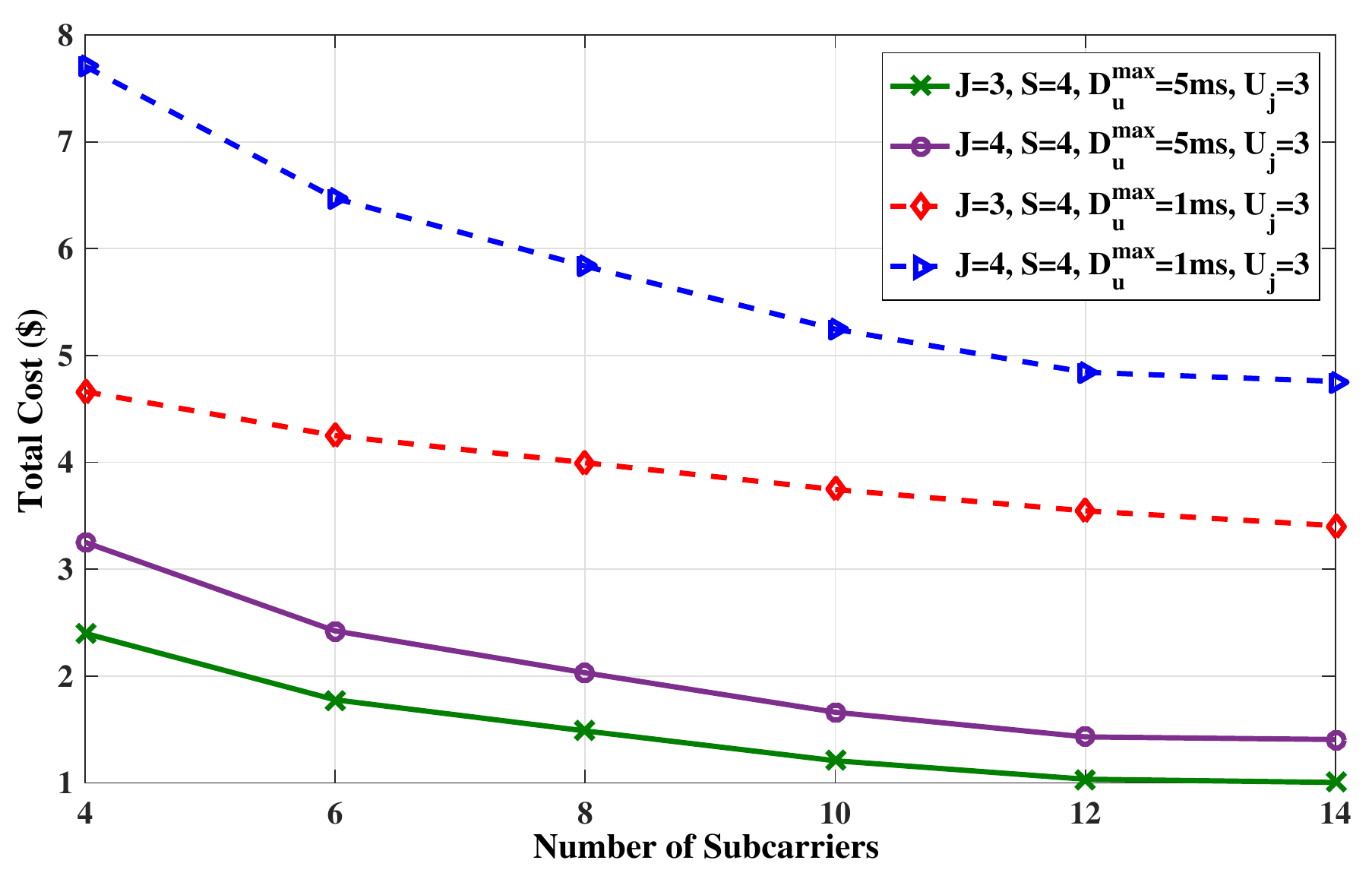}}
	  	\caption{ Total cost versus number of subcarriers  }
	  	\label{Fig2}
	  \end{figure}

Fig. \ref{Fig2} \textcolor{black}{illustrates} the total cost versus the number of subcarriers. As can be seen, the total cost decreases by increasing the number of subcarriers. This is due to the fact that increasing the number of subcarriers, increases the diversity gain of the network \textcolor{black}{which depends on channel model and frequency selectivity, i.e. the power delay profile. Therefore, the network has a higher degree of freedom in the subcarrier assignment and for each user, it can choose better subcarriers and in the same condition, less transmit power is needed. As a result, the total cost is reduced.}
	  
\begin{figure}[t]
	   	\centering
	   	\centerline{\includegraphics[width=0.53\textwidth]{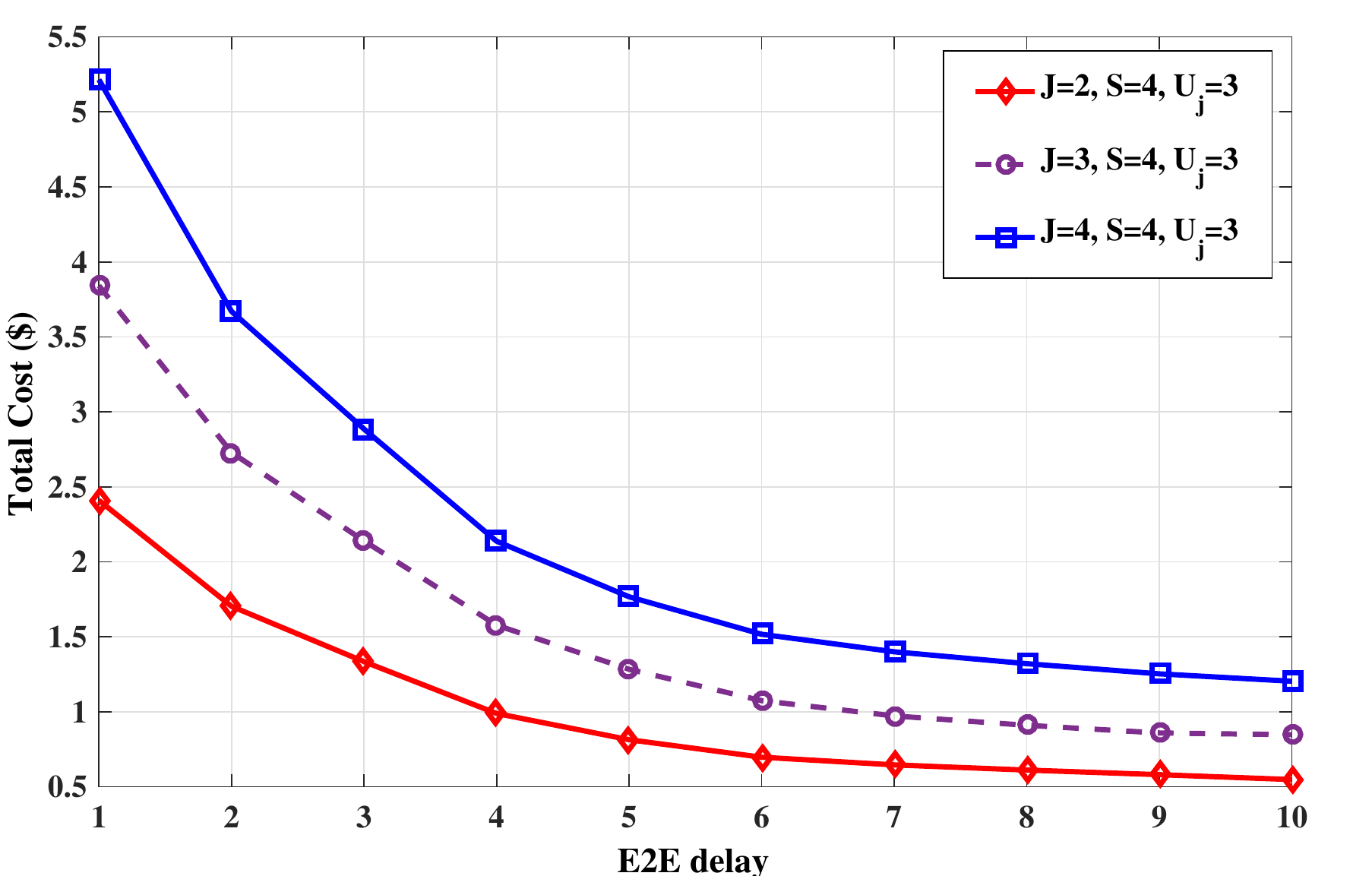}}
	   	\caption{ Total cost versus E2E delay for different number of BSs}
	   	\label{Fig3}
	   \end{figure}

In Fig. \ref{Fig3}, the effect of E2E delay of services on the total cost is addressed. In this regard, we assume all users request only one type of service with specific E2E delay requirement. Based on  C11 and C12, by increasing the delay, the required rate is decreased, hence the total transmit power is reduced.
	 Therefore, as can be seen from Fig. \ref{Fig3}, the total cost decreases by increasing E2E delay of services.

	   \subsection{Comparison between Joint Approach (JA) and Separate Approach (SA)}
	    In order to compare the proposed approach, i.e., JA, with the case where NFV-RA and R-RA are treated independently, i.e., SA, we assume a setting in which  we decompose NFV-RA and R-RA and solve them separately.  
   In R-RA, we set NFV delay manually in constraints  C6  to  fixed amount, e.g, $D_{u^s_j}^{\text{NFs}}=0.5$~ms. The new problem associated to R-RA in the SA can be written as
	\begin{align}\label{eqoa11}
	&\mathop {\min }\limits_{\scriptstyle{\bf{ P}},{\bf{T,X}},\hfill\atop
	{\bf{D}}\hfill}\sum_{j \in \mathcal{J}}\bigg(\sum_{s \in \mathcal{S}} \sum_{{u^s_j} \in \mathcal{U}^s_j}\sum_{k \in \mathcal{K}}x_{u^s_j}^{k} p_{u^s_j}^{k}+ \sum_{o_j \in \mathcal{O}_j} \sum_{l \in \mathcal{L}} \tau_{o_j}^{l} \hat p_{o_j}^{l}\bigg)\nonumber
	\\\text{s.t.:}&\text{~(C1)-(C4), (C6)-(C8), (C11)-(C12)}.
	\end{align}
		   \begin{figure}[t]
		   	\centering
		   	\centerline{\includegraphics[width=0.53\textwidth]{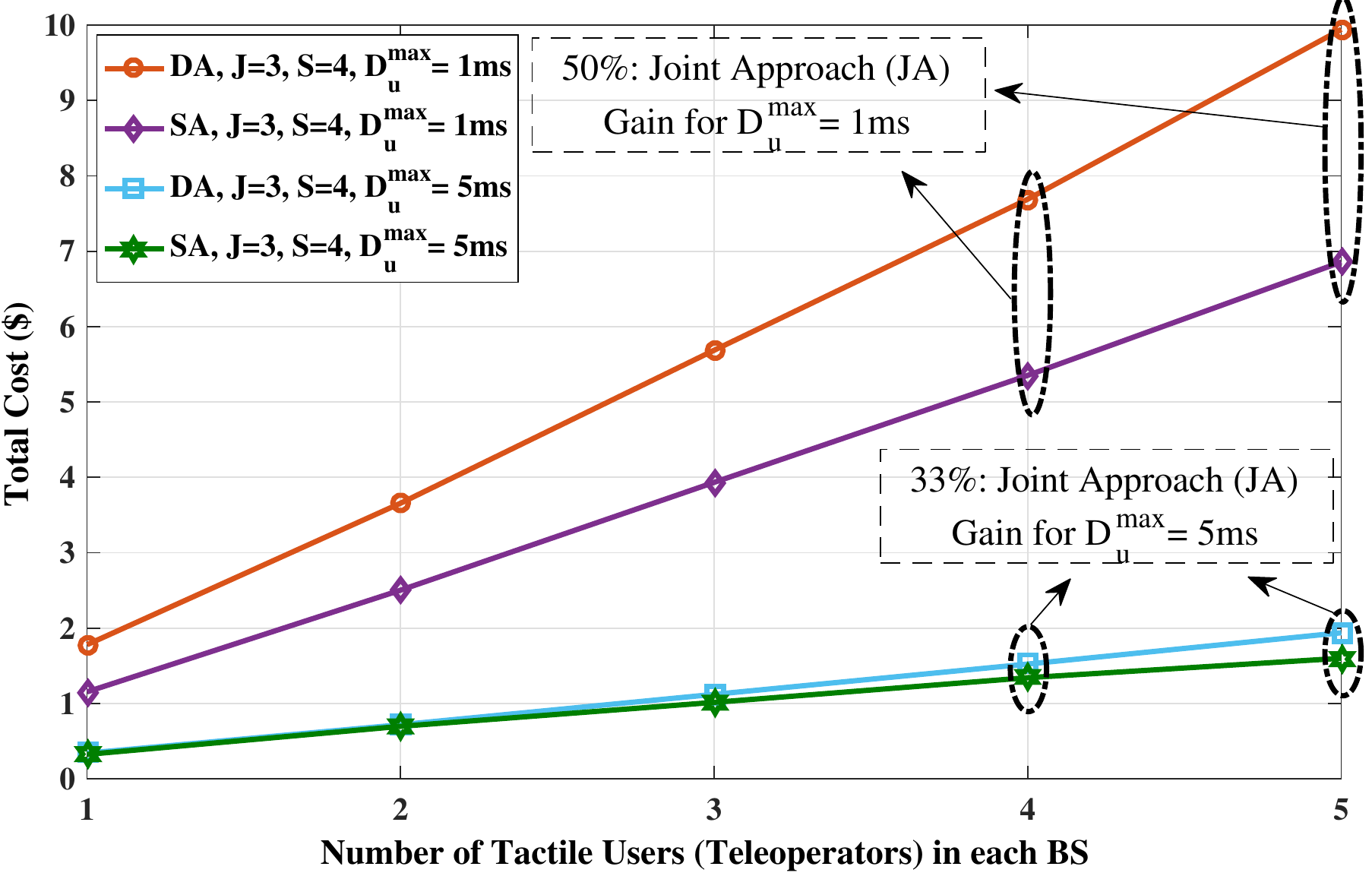}}
		   	\caption{ Total cost versus number of users in each BS for joint approach and separate approach }
		   	\label{Fig4}
		   \end{figure} 
	This new optimization problem can be solved with the ASM and SCA methods as well as the CVX toolbox. Similarly, the new problem for NFV-RA is as follows:
		\begin{equation}
	\begin{split}\label{NFVP-1}
	&\min_{{\bf{W}},{\bf{A}}} \varrho_2 \bigg( \sum_{j \in \mathcal{J}} \sum_{s \in \mathcal{S}} \sum_{{u^s_j} \in \mathcal{U}^s_j} \sum_{n_2 \in \mathcal{J}}  \max_{u^b_m,v,n_1}\{A_{u^s_j,f,n_2}^{u^b_m,v,n_1}\} \beta_{n_2}^{f}\frac{C_{u^s_j}}{\varOmega_{n_2}} \bigg) \\
	\text{s.t.:}& \text{\text{ C5, C9-C10. }}	
	\end{split}
	\end{equation}
		The NFV-RA problem can be solved with a heuristic Algorithm \ref{ALGNFV}.
	Now we set the E2E delay requirements to 5 ms and 1 ms and compare the cost associated to SA and JA. For both cases in SA, the delay associated to the NFV problem is set to 0.5 ms. As can be seen,
	there is not much difference between the coast associated to JA and SA for E2E delay of 5 ms. The reason is that the constraint for the R-RA case is only slightly modified compared to the JA. For E2E delay of 1 ms, however, we observe cost increase as high as 50\% when we choose SA over JA and the difference grows linearly when the number of tactile users.\footnote{This result is obtained based on the assumption that the ratio of $\varrho_2$ to $\varrho_1$ is 1. The realistic values for these 2 parameters depend on the actual network implementation. However, our simulation shows that the advantage of JA over SA remains significant for other values of $\varrho_2/\varrho_1$.} 
	
\section{Conclusion}\label{conclusion}
In this paper, we considered a heterogeneous wireless network with a variety of NSs for the Tactile Internet where  we considered queuing delays, transmission delays, and delays resulting from VNF execution
We then considered the joint radio resource allocation and NFV resource allocation to minimize the total cost function subject to guaranteeing E2E delay of each user. 
To solve this non-convex proposed resource allocation problem, we applied an ASM algorithm with SCA method.
Simulation results reveal that for smaller values of the E2E delay (e.g. 1 ms), the proposed joint scheme can significantly reduce the network costs compared to the case where the two problems are optimized separately.




%

\bibliographystyle{ieeetran}
\bibliography{Tactile}		

\begin{thebibliography}{10}
\providecommand{\url}[1]{#1}
\csname url@samestyle\endcsname
\providecommand{\newblock}{\relax}
\providecommand{\bibinfo}[2]{#2}
\providecommand{\BIBentrySTDinterwordspacing}{\spaceskip=0pt\relax}
\providecommand{\BIBentryALTinterwordstretchfactor}{4}
\providecommand{\BIBentryALTinterwordspacing}{\spaceskip=\fontdimen2\font plus
\BIBentryALTinterwordstretchfactor\fontdimen3\font minus
  \fontdimen4\font\relax}
\providecommand{\BIBforeignlanguage}[2]{{%
\expandafter\ifx\csname l@#1\endcsname\relax
\typeout{** WARNING: IEEEtran.bst: No hyphenation pattern has been}%
\typeout{** loaded for the language `#1'. Using the pattern for}%
\typeout{** the default language instead.}%
\else
\language=\csname l@#1\endcsname
\fi
#2}}
\providecommand{\BIBdecl}{\relax}
\BIBdecl

\bibitem{8403963}
H.~Ji, S.~Park, J.~Yeo, Y.~Kim, J.~Lee, and B.~Shim, ``Ultra-reliable and
  low-latency communications in {5G} downlink: Physical layer aspects,''
  \emph{IEEE Wireless Communications}, vol.~25, no.~3, pp. 124--130, JUNE 2018.

\bibitem{popovski20185g}
P.~Popovski, K.~F. Trillingsgaard, O.~Simeone, and G.~Durisi, ``{5G} wireless
  network slicing for {eMBB, URLLC, and mMTC}: A communication-theoretic
  view,'' \emph{arXiv preprint arXiv:1804.05057}, 2018.

\bibitem{she2016ensuring}
C.~She and C.~Yang, ``Ensuring the quality-of-service of {Tactile Internet},''
  in \emph{Proc Vehicular Technology Conference (VTC Spring)}, May 2016, pp.
  1--5.

\bibitem{simsek20165g}
M.~Simsek, A.~Aijaz, M.~Dohler, J.~Sachs, and G.~Fettweis, ``{5G-Enabled
  Tactile Internet},'' \emph{IEEE Journal on Selected Areas in Communications},
  vol.~34, no.~3, pp. 460--473, March 2016.

\bibitem{fettweis2014tactile}
G.~P. Fettweis, ``The {Tactile Internet}: Applications and challenges,''
  \emph{IEEE Vehicular Technology Magazine}, vol.~9, no.~1, pp. 64--70, March
  2014.

\bibitem{steinbach2012haptic}
E.~Steinbach, S.~Hirche, M.~Ernst, F.~Brandi, R.~Chaudhari, J.~Kammerl, and
  I.~Vittorias, ``Haptic communications,'' \emph{Proceedings of the IEEE}, vol.
  100, no.~4, pp. 937--956, April 2012.

\bibitem{han2015network}
B.~Han, V.~Gopalakrishnan, L.~Ji, and S.~Lee, ``Network function
  virtualization: Challenges and opportunities for innovations,'' \emph{IEEE
  Communications Magazine}, vol.~53, no.~2, pp. 90--97, Feb 2015.

\bibitem{mijumbi2016network}
R.~Mijumbi, J.~Serrat, J.~Gorricho, N.~Bouten, F.~D. Turck, and R.~Boutaba,
  ``Network function virtualization: State-of-the-art and research
  challenges,'' \emph{IEEE Communications Surveys Tutorials}, vol.~18, no.~1,
  pp. 236--262, Firstquarter 2016.

\bibitem{8115155}
M.~{Moltafet}, N.~{Mokari}, M.~R. {Javan}, H.~{Saeedi}, and H.~{Pishro-Nik},
  ``A new multiple access technique for {5G}: Power domain sparse code multiple
  access ({PSMA}),'' \emph{IEEE Access}, vol.~6, pp. 747--759, 2018.

\bibitem{martins2014clickos}
J.~Martins, M.~Ahmed, C.~Raiciu, V.~Olteanu, M.~Honda, R.~Bifulco, and
  F.~Huici, ``Clickos and the art of network function virtualization,'' in
  \emph{Proc Proceedings of the 11th USENIX Conference on Networked Systems
  Design and Implementation}.\hskip 1em plus 0.5em minus 0.4em\relax USENIX
  Association, 2014, pp. 459--473.

\bibitem{7534741}
J.~G. Herrera and J.~F. Botero, ``Resource allocation in {NFV}: A comprehensive
  survey,'' \emph{IEEE Transactions on Network and Service Management},
  vol.~13, no.~3, pp. 518--532, Sept 2016.

\bibitem{7243304}
R.~Mijumbi, J.~Serrat, J.~Gorricho, N.~Bouten, F.~D. Turck, and R.~Boutaba,
  ``Network function virtualization: State-of-the-art and research
  challenges,'' \emph{IEEE Communications Surveys Tutorials}, vol.~18, no.~1,
  pp. 236--262, Firstquarter 2016.

\bibitem{riera2014virtual}
J.~F. Riera, E.~Escalona, J.~Batalle, E.~Grasa, and J.~A. Garcia-Espin,
  ``Virtual network function scheduling: Concept and challenges,'' in
  \emph{Proc Smart Communications in Network Technologies (SaCoNeT), 2014
  International Conference on}.\hskip 1em plus 0.5em minus 0.4em\relax IEEE,
  2014, pp. 1--5.

\bibitem{cao2017vnf}
J.~Cao, Y.~Zhang, W.~An, X.~Chen, J.~Sun, and Y.~Han, ``{VNF-FG} design and
  {VNF} placement for {5G} mobile networks,'' \emph{Science China Information
  Sciences}, vol.~60, no.~4, p. 040302, 2017.

\bibitem{monteleone2013session}
G.~Monteleone and P.~Paglierani, ``Session border controller virtualization
  towards" service-defined" networks based on {NFV} and {SDN},'' in \emph{Proc
  2013 IEEE SDN for Future Networks and Services (SDN4FNS)}.\hskip 1em plus
  0.5em minus 0.4em\relax IEEE, 2013, pp. 1--7.

\bibitem{riggio2016scheduling}
R.~Riggio, A.~Bradai, D.~Harutyunyan, T.~Rasheed, and T.~Ahmed, ``Scheduling
  wireless virtual networks functions.'' \emph{IEEE Trans. Network and Service
  Management}, vol.~13, no.~2, pp. 240--252, 2016.

\bibitem{mijumbi2015design}
R.~Mijumbi, J.~Serrat, J.-L. Gorricho, N.~Bouten, F.~De~Turck, and S.~Davy,
  ``Design and evaluation of algorithms for mapping and scheduling of virtual
  network functions,'' in \emph{Proc Network Softwarization (NetSoft), 2015 1st
  IEEE Conference on}.\hskip 1em plus 0.5em minus 0.4em\relax IEEE, 2015, pp.
  1--9.

\bibitem{beck2015coordinated}
M.~T. Beck and J.~F. Botero, ``Coordinated allocation of service function
  chains,'' in \emph{Proc Global Communications Conference (GLOBECOM), 2015
  IEEE}.\hskip 1em plus 0.5em minus 0.4em\relax IEEE, 2015, pp. 1--6.

\bibitem{wang2016joint}
L.~Wang, Z.~Lu, X.~Wen, R.~Knopp, and R.~Gupta, ``Joint optimization of service
  function chaining and resource allocation in network function
  virtualization,'' \emph{IEEE Access}, vol.~4, pp. 8084--8094, 2016.

\bibitem{she7636814tactile}
C.~She and C.~Yang, ``Energy efficient design for tactile internet,'' in
  \emph{2016 IEEE/CIC International Conference on Communications in China
  (ICCC)}, July 2016, pp. 1--6.

\bibitem{she2016uplink}
C.~She, C.~Yang, and T.~Q. Quek, ``Uplink transmission design with massive
  machine type devices in {Tactile Internet},'' in \emph{Proc Globecom
  Workshops (GC Wkshps), 2016 IEEE}.\hskip 1em plus 0.5em minus 0.4em\relax
  IEEE, 2016, pp. 1--6.

\bibitem{aijaz2016towards}
A.~Aijaz, ``Towards {5G-enabled Tactile Internet}: Radio resource allocation
  for haptic communications,'' in \emph{Proc Wireless Communications and
  Networking Conference Workshops (WCNCW), 2016 IEEE}.\hskip 1em plus 0.5em
  minus 0.4em\relax IEEE, 2016, pp. 145--150.

\bibitem{gholipoor2018cross}
N.~{Gholipoor}, H.~{Saeedi}, and N.~{Mokari}, ``Cross-layer resource allocation
  for mixed tactile internet and traditional data in {SCMA} based wireless
  networks,'' in \emph{2018 IEEE Wireless Communications and Networking
  Conference Workshops (WCNCW)}, April 2018, pp. 356--361.

\bibitem{gholipoor2019cloud}
N.~Gholipoor, S.~Parsaeefard, M.~R. Javan, N.~Mokari, and H.~Saeedi,
  ``Cloud-based queuing model for tactile internet in next generation of
  {RAN},'' \emph{arXiv preprint arXiv:1901.09389}, 2019.

\bibitem{gholipoor2019resource}
N.~Gholipoor, S.~Parsaeefard, M.~R. Javan, N.~Mokari, H.~Saeedi, and
  H.~Pishro-Nik, ``Resource management and admission control for tactile
  internet in next generation of {RAN},'' \emph{arXiv preprint
  arXiv:1907.01403}, 2019.

\bibitem{roos2006broadband}
A.~Roos, A.~T. Schwarzbacher, and S.~Wieland, ``Broadband wireless internet
  access in public transportation,'' in \emph{Proc. VDE Kongress-Innovations in
  {Europe, Aachen, Germany}}, vol.~1, 2006, pp. 65--70.

\bibitem{7497015}
Y.~{Niu}, C.~{Gao}, Y.~{Li}, L.~{Su}, D.~{Jin}, Y.~{Zhu}, and D.~O. {Wu},
  ``Energy-efficient scheduling for mmwave backhauling of small cells in
  heterogeneous cellular networks,'' \emph{IEEE Transactions on Vehicular
  Technology}, vol.~66, no.~3, pp. 2674--2687, March 2017.

\bibitem{liao2013performance}
R.~Liao, B.~Bellalta, J.~Barcelo, V.~Valls, and M.~Oliver, ``Performance
  analysis of {IEEE 802.11 ac} wireless backhaul networks in saturated
  conditions,'' \emph{EURASIP Journal on Wireless Communications and
  Networking}, vol. 2013, no.~1, p. 226, 2013.

\bibitem{4224302}
K.~{Karakayali}, J.~H. {Kang}, M.~{Kodialam}, and K.~{Balachandran},
  ``Cross-layer optimization for {OFDMA-Based} wireless mesh backhaul
  networks,'' in \emph{2007 IEEE Wireless Communications and Networking
  Conference}, March 2007, pp. 276--281.

\bibitem{8651725}
M.~{Nakamura}, G.~K. {Tran}, and K.~{Sakaguchi}, ``Interference management for
  millimeter-wave mesh backhaul networks,'' in \emph{2019 16th IEEE Annual
  Consumer Communications Networking Conference (CCNC)}, Jan 2019, pp. 1--4.

\bibitem{7452271}
S.~{Abdelwahab}, B.~{Hamdaoui}, M.~{Guizani}, and T.~{Znati}, ``Network
  function virtualization in {5G},'' \emph{IEEE Communications Magazine},
  vol.~54, no.~4, pp. 84--91, April 2016.

\bibitem{6553675}
X.~{Costa-Perez}, J.~{Swetina}, T.~{Guo}, R.~{Mahindra}, and S.~{Rangarajan},
  ``Radio access network virtualization for future mobile carrier networks,''
  \emph{IEEE Communications Magazine}, vol.~51, no.~7, pp. 27--35, July 2013.

\bibitem{filippou2019flexible}
M.~C. Filippou, D.~Sabella, and V.~Riccobene, ``Flexible{MEC} service
  consumption through edge host zoning in {5G} networks,'' \emph{arXiv preprint
  arXiv:1903.01794}, 2019.

\bibitem{8533343}
T.~X. {Tran} and D.~{Pompili}, ``Joint task offloading and resource allocation
  for multi-server mobile-edge computing networks,'' \emph{IEEE Transactions on
  Vehicular Technology}, vol.~68, no.~1, pp. 856--868, Jan 2019.

\bibitem{8436039}
H.~{Guo}, J.~{Liu}, and J.~{Zhang}, ``Computation offloading for multi-access
  mobile edge computing in ultra-dense networks,'' \emph{IEEE Communications
  Magazine}, vol.~56, no.~8, pp. 14--19, August 2018.

\bibitem{chang1995effective}
C.-S. Chang and J.~A. Thomas, ``Effective bandwidth in high-speed digital
  networks,'' \emph{IEEE Journal on Selected Areas in Communications}, vol.~13,
  no.~6, pp. 1091--1100, August 1995.

\bibitem{4543084}
J.~{Tang} and X.~{Zhang}, ``Cross-layer-model based adaptive resource
  allocation for statistical {QoS} guarantees in mobile wireless networks,''
  \emph{IEEE Transactions on Wireless Communications}, vol.~7, no.~6, pp.
  2318--2328, June 2008.

\bibitem{1210731}
and R.~{Negi}, ``Effective capacity: a wireless link model for support of
  quality of service,'' \emph{IEEE Transactions on Wireless Communications},
  vol.~2, no.~4, pp. 630--643, July 2003.

\bibitem{bertsekas1997nonlinear}
D.~P. Bertsekas, ``Nonlinear programming,'' \emph{Journal of the Operational
  Research Society}, vol.~48, no.~3, pp. 334--334, 1997.

\bibitem{wang2011iterative}
T.~Wang and L.~Vandendorpe, ``Iterative resource allocation for maximizing
  weighted sum min-rate in downlink cellular {OFDMA} systems,'' \emph{IEEE
  Transactions on Signal Processing}, vol.~59, no.~1, pp. 223--234, 2011.

\bibitem{6678362}
D.~T. {Ngo}, S.~{Khakurel}, and T.~{Le-Ngoc}, ``Joint subchannel assignment and
  power allocation for {OFDMA} femtocell networks,'' \emph{IEEE Transactions on
  Wireless Communications}, vol.~13, no.~1, pp. 342--355, January 2014.

\bibitem{mokdad2016radio}
A.~Mokdad, P.~Azmi, and N.~Mokari, ``Radio resource allocation for
  heterogeneous traffic in {GFDM-NOMA} heterogeneous cellular networks,''
  \emph{IET Communications}, vol.~10, no.~12, pp. 1444--1455, 2016.

\bibitem{rezvani2019fairness}
S.~Rezvani, N.~Mokari, M.~R. Javan, and E.~Jorswieck, ``Fairness and
  transmission-aware caching and delivery policies in {OFDMA-based HetNets},''
  \emph{IEEE Transactions on Mobile Computing}, 2019.

\bibitem{moltafet2018optimal}
M.~Moltafet, P.~Azmi, N.~Mokari, M.~R. Javan, and A.~Mokdad, ``Optimal and fair
  energy efficient resource allocation for energy
  harvesting-enabled-{PD-NOMA}-based {HetNets},'' \emph{IEEE Transactions on
  Wireless Communications}, vol.~17, no.~3, pp. 2054--2067, 2018.

\bibitem{ngo2014joint}
D.~T. Ngo, S.~Khakurel, and T.~Le-Ngoc, ``Joint subchannel assignment and power
  allocation for {OFDMA} femtocell networks,'' \emph{IEEE Transactions on
  Wireless Communications}, vol.~13, no.~1, pp. 342--355, January 2014.

\end{thebibliography}
\bibliographystyle{ieeetr}
\end{document}